\documentclass{egpubl}
\usepackage{eg2024}

\ConferencePaper        %

\CGFccbync

\usepackage[T1]{fontenc}
\usepackage{dfadobe}  

\usepackage{cite}  %
\BibtexOrBiblatex
\electronicVersion
\PrintedOrElectronic
\ifpdf \usepackage[pdftex]{graphicx} \pdfcompresslevel=9
\else \usepackage[dvips]{graphicx} \fi

\usepackage{egweblnk} 
\usepackage{fancybox}
\usepackage{tikz}
\usepackage{pgfplots}
\usepackage{overpic}
\usepackage{xcolor}

\usepackage{wrapfig}
\usepackage{floatflt}
\usepgfplotslibrary{fillbetween}

\usepackage[normalem]{ulem}
\usepackage{amsmath,amssymb}
\usepackage[ruled,linesnumbered]{algorithm2e}
\usepackage{graphicx}
\usepackage{subfigure}
\usepackage{todonotes}

\newcommand{\dav}[1]{#1}
\newcommand{\batou}[1]{#1}

\newcommand{\projmu}[1][\theta]{P^{#1}_\# \mu}
\newcommand{\projnu}[1][\theta]{P^{#1}_\# \nu}
\newcommand{\Hy}{\mathbb{H}}
\newcommand{\Sp}{\mathbb{S}}
\newcommand{\Pp}{\mathbb{P}}
\newcommand{\Span}{\text{span}}
\newcommand{\Euc}{\mathbb{R}}
\newcommand{\normalize}[1]{\frac{#1}{\|#1\|}}
\newcommand{\dotE}[2]{\langle #1,#2 \rangle}
\newcommand{\dotL}[2]{\langle #1,#2 \rangle_{\mathbb{L}}}
\newcommand{\icdf}[1]{F_{#1}^{-1}}
\newcommand{\acos}{\text{arccos}}
\newcommand{\acosh}{\text{arccosh}}

\newcommand{\sgn}{\text{sign}}
\newcommand{\Log}{\text{Log}}
\newcommand{\Exp}{\text{Exp}}
\newcommand{\atan}{\text{arctan2}}
\newcommand{\mesh}{\mathcal{M}}

\newcommand{\xp}{\mathbf{x}}
\newcommand{\yp}{\mathbf{y}}
\newcommand{\wpp}{\mathbf{w}}
\newcommand{\ep}{\mathbf{e}}
\newcommand{\dpp}{\mathbf{d}}
\newcommand{\xO}{\mathbf{x}_O}
\newcommand{\thetap}{\mathbf{\theta}}
\newcommand{\vp}{\mathbf{v}}
\newcommand{\gp}{\mathbf{g}}
\newcommand{\qp}{\mathbf{q}}

\title{Non-Euclidean Sliced Optimal Transport Sampling}

\author[B. Genest, N. Courty \& D. Coeurjolly]
 {\parbox{\textwidth}{\centering Baptiste Genest$^{1}$\orcid{0009-0009-7718-5553},
        Nicolas Courty$^2$\orcid{0000-0003-1353-0126}
         and David Coeurjolly$^{1}$\orcid{0000−0003−3164−8697} 
      }
         \\
 {\parbox{\textwidth}{\centering $^1$Univ Lyon, CNRS, Lyon1, INSA, LIRIS, France\\
          $^2$Université Bretagne Sud, IRISA, CNRS, France
        }
 }
 }

\begin{document}

\teaser{
  \centering\includegraphics[width=.95\textwidth]{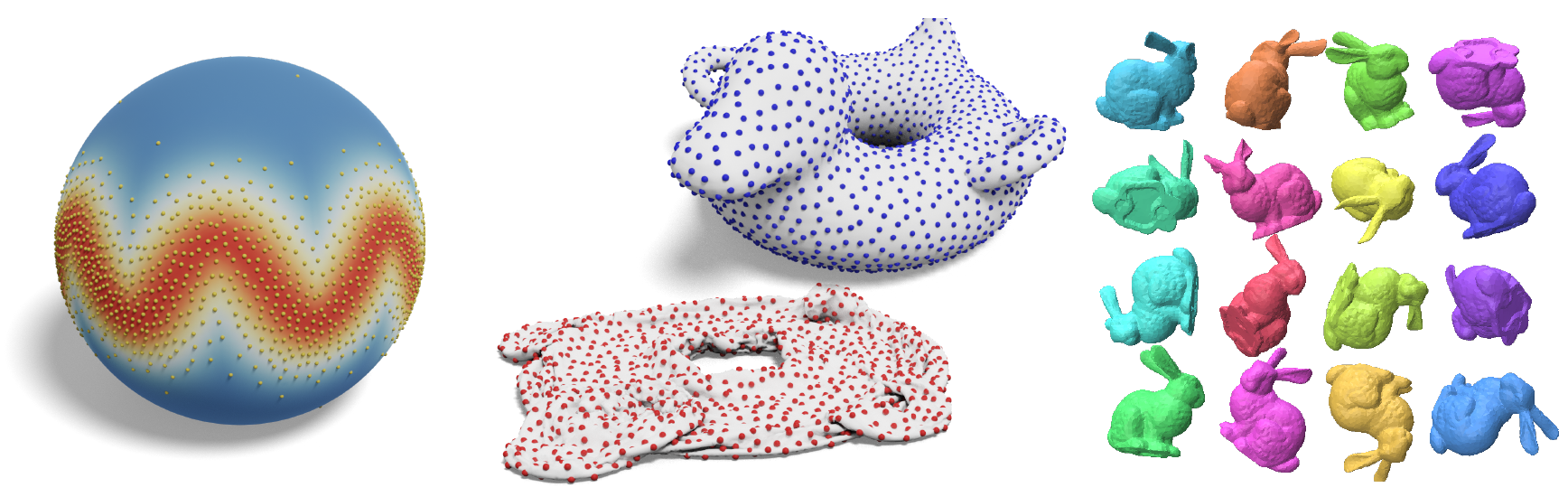}
 
  \caption{We propose a new technique to generate well-dispersed samples on non-Euclidean domains (spherical,  hyperbolic and projective spaces) using an extension of the sliced optimal transport sampling. As an example, this allows us to sample probability measures on the high-dimensional sphere \emph{(left)}. Using the uniformization theorem to conformally embed discrete manifolds to spherical or hyperbolic spaces, we can also generate blue noise samples in a purely intrinsic manner (red samples on the flatten geometry that exhibits blue noise properties when mapped back to a better embedding in $\mathbb{R}^3$ in blue). Finally, we also demonstrate that such an approach can be used to \emph{blue noise} sample unit quaternions (hence rotations) on the projective space of dimension 3 (right).\label{fig:teaser}}
}

\maketitle

\begin{abstract}
    In machine learning and computer graphics, a fundamental task is the approximation of a probability density 
    function through a well-dispersed collection of samples. Providing a formal metric for measuring the distance 
    between probability measures on general spaces, Optimal Transport (OT) emerges as a pivotal theoretical framework 
    within this context. However, the associated computational burden is prohibitive in most real-world scenarios.
     Leveraging the simple structure of OT in $1D$, Sliced Optimal Transport (SOT) has appeared as an efficient alternative 
     to generate samples in Euclidean spaces. This paper pushes the boundaries of SOT utilization in computational geometry
      problems by extending its application to sample densities residing on more diverse mathematical domains, including the 
      spherical space $\mathbb{S}^d$, the hyperbolic plane $\mathbb{H}^d$, and the real projective plane $\mathbb{P}^d$. 
      Moreover, it ensures the quality of these samples by achieving a blue noise characteristic, regardless of the dimensionality
       involved. The robustness of our approach is highlighted through its application to various geometry processing tasks, 
       such as the intrinsic blue noise sampling of meshes, as well as the sampling of directions and rotations. 
       These applications collectively underscore the efficacy of our methodology.


\end{abstract}  

\section{Introduction}
\label{sec:introduction}

In recent years, Optimal Transport has become a key mathematical framework for manipulating
generalized probability density functions (\emph{e.g.} \cite{villani2009optimal}). The most general way 
to describe the interest of OT is that it allows quantifying meaningfully how costly
it is to move masses from a generalized probability density function to another one. This defines
a natural notion of distance between probability measures, the Wasserstein distance, allowing the design of displacement 
interpolations between measures or when dealing with more than two measures, the notion of Wasserstein
barycenter.

The high versatility of the framework and the numerous developments of efficient numerical
 solvers make the OT become standard in many machine learning \cite{huang2016supervised,courty2016optimal,arjovsky2017wasserstein},%
 computer vision, or computer graphics applications  
 \cite{de2012blue,solomon2014earth,solomon2015convolutional,bonneel2015sliced,qin2017wasserstein,nader2018instant,BC19,paulin2020sliced,salaun2022scalable} 
 (see \cite{bonneel2023survey} for a recent survey).

 Among computer graphics applications, OT has become a widely spread tool for  point pattern design and Monte
  Carlo integration \cite{qin2017wasserstein,paulin2020sliced,salaun2022scalable}. The main argument is that OT offers a 
  mathematical framework to characterize well-distributed, or blue noise, samples in a domain leading to an efficient Monte Carlo integration or signal reconstruction \cite{singh2019analysis}.
   This can be achieved by optimizing the samples positions such that the Wasserstein distance to
  the uniform measure in the domain is minimized.
 More recently, OT on non-Euclidean spaces has been developed  in the machine learning context, as  
 it allows efficiently processing of data for which a spherical or hyperbolic geometry is a natural representation space~\cite{bonet2022spherical,bonet2022hyperbolic}. In geometry processing, a spherical or hyperbolic
  embedding of geometrical objects can be at the core of many surface parametrization, texture mapping or shape matching problems 
  \cite{haker2000conformal,gu2003global,gotsman2003fundamentals,kharevych2006discrete,crane2013robust,baden2018mobius,schmidt2020inter,Gillespie:2021:DCE}.
The challenge addressed in this paper is the design of an OT driven sampling techniques on Riemannian manifolds with applications to computer graphics.%

\paragraph*{Contributions.}  Relying on sliced optimal transport formulation for the sphere and the hyperbolic space formulated \dav{by Bonet et al.} \cite{bonet2022spherical,bonet2022hyperbolic}, we propose a blue noise sampling strategy of probability measures on these
non-Euclidean spaces. This is achieved by providing explicit formulas for the samples advection steps
and direction pooling in a Riemannian gradient descent approach. We then demonstrate the strength of the approach
to efficiently sample meshes through the uniformization theorem allowing transforming the intrinsic blue noise
sampling problem on the mesh, to a blue noise sampling problem in $\Sp^2$ or $\Hy^2$ depending on the mesh topology. We also highlight 
the interest of the approach \batou{ through projective plane sampling that can be used to sample 3D rotations (by sampling quaternions in 4d),
 as well as various geometric objects befined by projective equations (\emph{e.g.} lines, directions...). }

\section{Background}

\paragraph*{Optimal transport.}   Given two measures $\mu$ and $\nu$, over some domain $\Omega$, and a function $c(x,y)$ that dictates the cost of moving a particle from
$x$ to $y$ in $\Omega$, one can define the Optimal Transport problem from $\mu$ to
$\nu$ as
\begin{equation}
\label{eq:OT}
\min_{\pi \in \Pi(\mu,\nu)} \int_{\Omega} c(x,y)d\pi(x,y)\,.
\end{equation}
where $\Pi(\mu,\nu)$ is the set of couplings:
\begin{align*}
    \{\pi \in \mathcal{P}(\Omega\times\Omega),& \forall A \subset \Omega, \pi(A \times \Omega) = \mu(A), \pi(\Omega\times A) = \nu(A) \}\,.
\end{align*}
In most contexts, $c(x,y) = d^p(x,y)$ where $d$ is a distance on
$\Omega$ (\emph{e.g.} \cite{peyre2019computational}). In such cases we call the minimum cost the $p-$Wasserstein
distance between $\mu$ and $\nu$, $W^p_p(\mu,\nu)$.  The interest of
using measures is that its general enough to handle both discrete and
continuous objects at the same time. Depending on the nature of the measures, discrete-to-discrete, 
semi-discrete, or continuous-to-continuous, a huge literature exists on numerical methods to 
efficiently solve OT problems \cite{peyre2019computational,flamary2021pot}.

\paragraph*{Sliced Optimal Transport.}  Among alternative numerical methods, we are interested in fast approximation techniques 
that scale up with the size of the discrete problem and the dimension. First, we observe that the one-dimensional OT problem admits the following closed form solution:
\begin{equation}
    W_p^p(\mu,\nu) = \int_0^1 |\icdf{\mu}(u) -\icdf{\nu}(u)|^p
    du\,,
\end{equation}
where $F_{\mu}$ is the cumulative function of the 1D density $\mu$, {and $\icdf{\mu}$ its generalized inverse, or quantile function}. For
$p=1$, {one can derive the equivalent formula:}
\begin{equation}
\label{eq:W1}
    W_1(\mu,\nu) = \int_0^1 |F_{\mu}(u) -F_{\nu}(u)| du\,.
\end{equation}
 The transport plan is then simply given by associating the ith point
 of $\mu$ to the ith point of $\nu$ (see for example \cite{peyre2019computational}) {in the case when $\mu$ and $\nu$ are both discrete with the same number of atoms}. The obtained result is the
 mapping that minimizes the cost to transport $\mu$ to $\nu$.  Hence,
 a very natural idea is to break a $d$ dimensional OT problem into an
 infinity of 1 dimensional one. Such an approach is referred to as
 \emph{Sliced Optimal Transport} since it amounts to projecting the
 measures onto 1D \emph{slices}
 \cite{Pitie2005,rabin2011wasserstein,bonneel2015sliced}. {Given
   a direction $\theta\in\Sp^{d-1}$ and the projection  $P^\theta(\xp):=\langle
   \xp,\theta\rangle$ of any , for all
   $\xp\in\mathbb{R}^{d}$, the sliced Wasserstein distance } is defined as
\begin{equation}
    SW_p^p(\mu,\nu)  := \int_{\mathbb{S}^{d-1}} W_p^p(\projmu,\projnu)\,\,d\lambda(\theta)\,,
\end{equation}
where $P_\#^\theta \mu$ is the image measure of $\mu$ by the
projection operator. The sliced approach  receives a lot of attention in the literature as it is topologically equivalent to OT  \cite{nadjahi2020statistical}  with bounded approximation of  $W_p$ \cite{bonnotte2013unidimensional}.
{On the
  algorithmic side, the integral over ${\mathbb{S}^{d-1}}$ is
  obtained used a Monte Carlo approach: we draw random directions
  uniformly on ${\mathbb{S}^{d-1}}$ and accumulate 1d Wasserstein distances. The
computational advantage is that each  1d slice $W_p^p$ only requires to
sort the points, leading to an overall computation cost in
$\mathcal{O}(K\cdot n( d+\log(n)))$ time complexity if $K$ denotes the
number of slices used in the Monte Carlo estimation.}

\begin{figure*}
  \centering
  ~~~~~~~~~~~~~~~~~~~~~~~~~\begin{tabular}{ccc}
    \begin{overpic}[width=4cm]{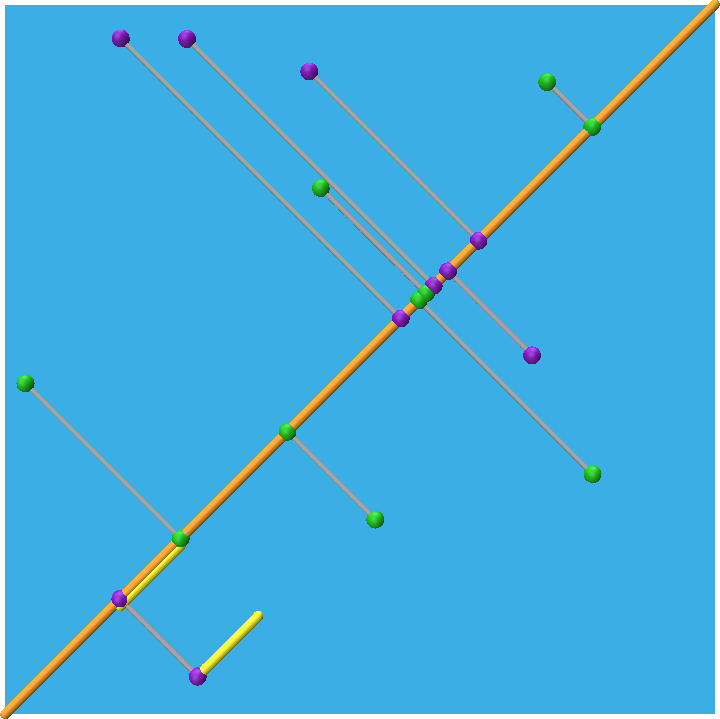}
      \put(-40,80){\includegraphics[width=.3cm]{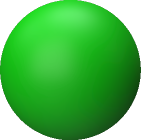}}
      \put(-25,82){$\mu$}
      \put(-40,70){\includegraphics[width=.3cm]{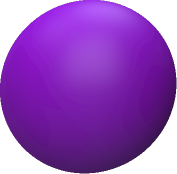}}
      \put(-25,72){$\nu$}
      \put(-55,55){\includegraphics[width=1cm,height=0.17cm]{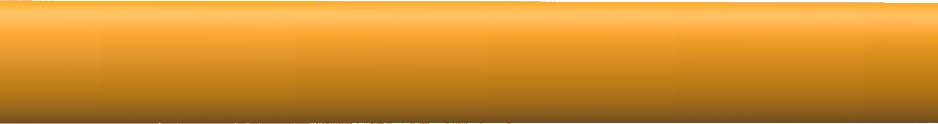}}
      \put(-25,55){$\theta$}
      \put(-55,45){\includegraphics[angle=270,width=1cm]{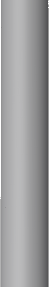}}
      \put(-25,41){$P^\theta(\xp)$}
      \put(-55,25){\includegraphics[width=1cm,height=0.17cm]{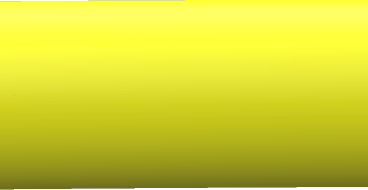}}
      \put(-25,24){$T(\xp)$}
      \put(75,2){\color{white}$\mathbb{R}^2$}
      \put(86,2){\includegraphics[width=0.5cm]{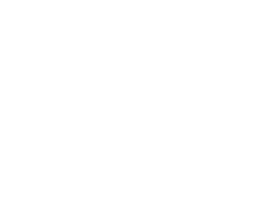}}
      \put(200,0){\includegraphics[width=0.5cm]{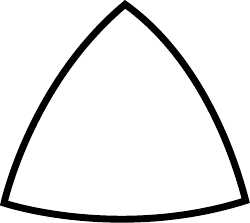}}
      \put(310,0){\includegraphics[width=0.5cm]{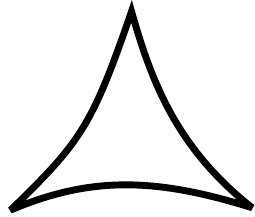}}
      \put(190,2){$\Sp^2$}
      \put(300,2){$\Hy^2$}

  \end{overpic}&
   \begin{overpic}[width=4cm]{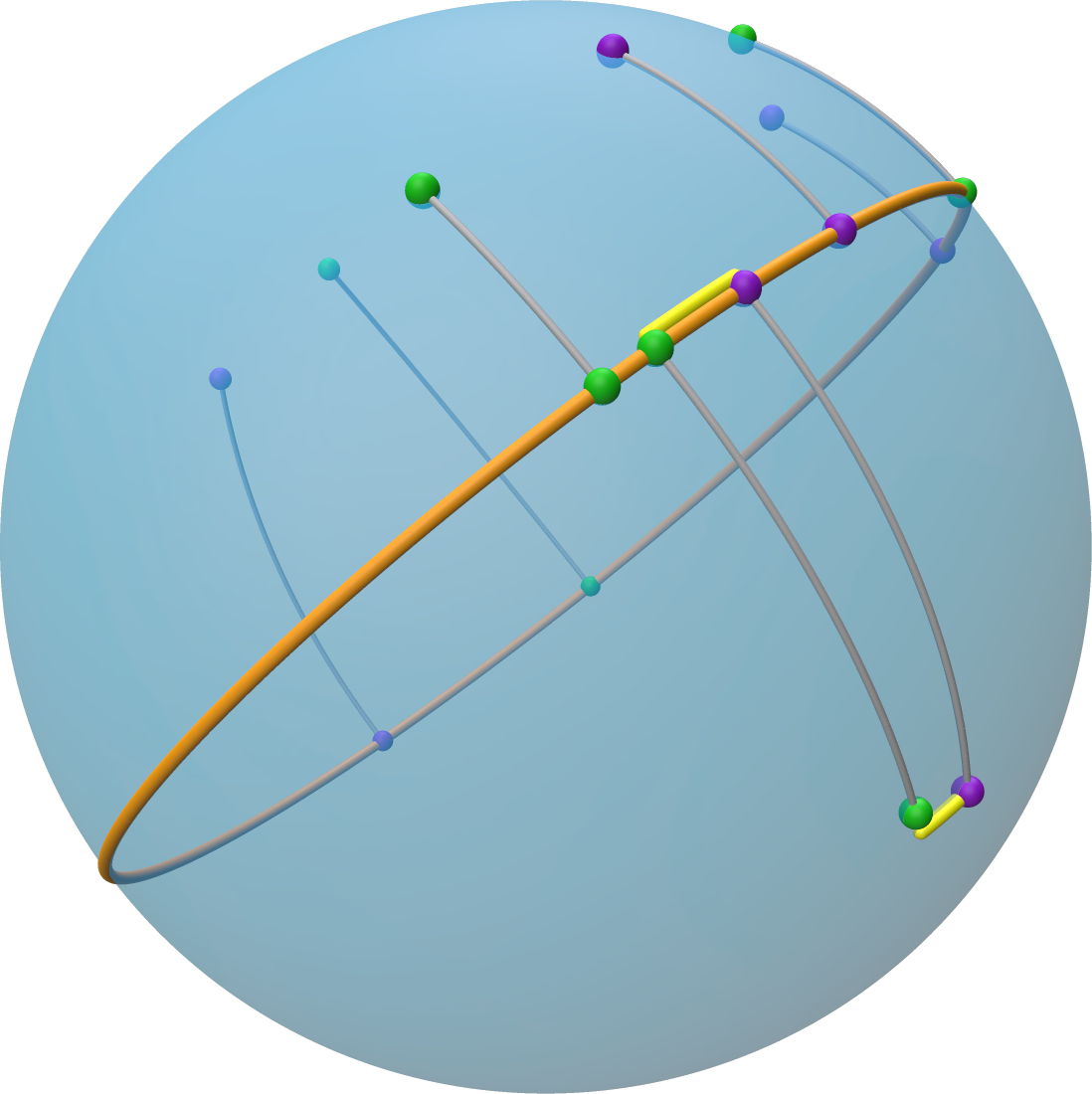}
   \end{overpic}
   &
   \begin{overpic}[width=4cm]{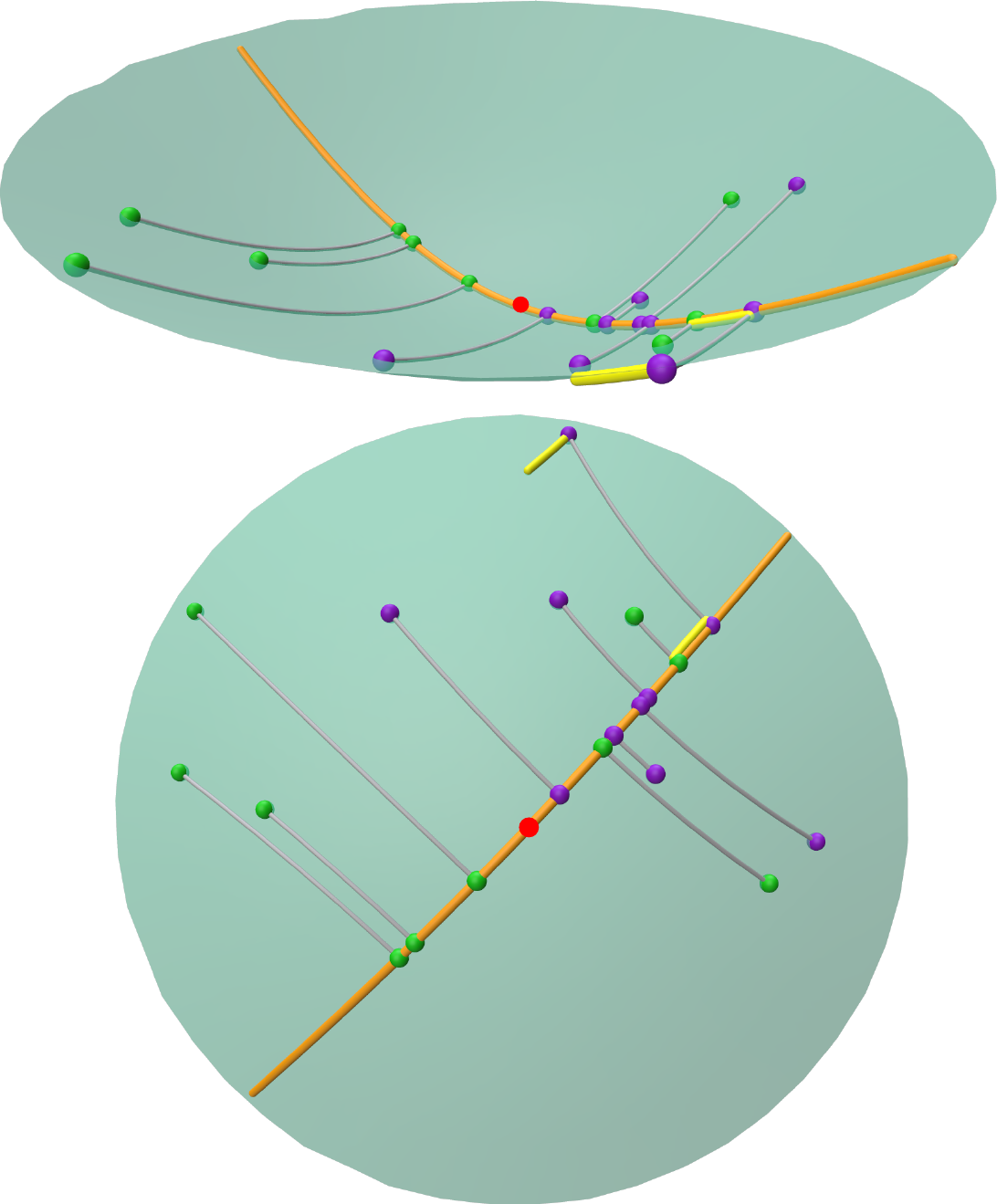}
   \end{overpic}
  \end{tabular}
\caption{\textbf{Sliced optimal transport sampling and notations}: from left to right, on the Euclidean domain (zero curvature metric space), 
on the spherical one (positive constant curvature metric space),
 and on the hyperbolic model (Lorentz's model with only a part of the hyperboloid, negative curvature metric space). We only illustrate the assignment and the associated advection
for a single sample (yellow bars).\label{fig:models}}
\end{figure*}

\paragraph*{Sliced Optimal Transport Sampling (SOTS).} {In the context of 
Monte Carlo sampling, Paulin et al. \cite{paulin2020sliced} leveraged the Euclidean sliced optimal transport
formulation to optimize a point set such that it better approximates
a given target distribution, in the sense of the $SW_2$ metric. In this Monte Carlo rendering setting,  given a target measure $\nu$ in $[0,1)^d$  (uniform measure for blue noise sampling),
  the objective is to construct $n$ samples $\{\xp_i\}\in[0,1)^d$ defining the discrete  distribution  $\mu = \sum_{i=1}^n
    \delta_{\xp_i}$, such that $SW_2(\mu,\nu)$ is minimized.}
    One iteration of the sliced optimal transport sampling, SOTS for short,
algorithm is the following, if $\mu = \sum_{i=1}^n \delta_{\xp_i}$ and
if $\nu$ is a continuous measure with closed form projection formula
on a line (mainly the uniform measure over a ball or a square), we
iterate:
\begin{equation}
    \xp^{(K+1)}_i = \xp^{(K)}_i + \frac{\gamma}{L} \sum_{l=1}^{L}\left(T_{l}\left(P^{\theta_l}(\xp^{(K)}_i)\right) - 
    P^{\theta_l}\left(\xp^{(K)}_i\right)\right)\,,
\end{equation}
where $T_l$ is the transport plan associated with the solution of the
continuous-to-discrete problem between $\projnu[\theta_l]$ and
$\projmu[\theta_l]$ and $\gamma > 0$ is a step
size (see Fig.\ref{fig:models}-left). For the sake of simplicity, the $P^\theta(\xp)$ notation refers
to the projection of the sample $\xp$ onto the slice $\theta$
(\emph{i.e.} $P_\#^\theta\mu = \sum_i \delta_{P^\theta(\xp_i)}$). Intuitively, we move
 each point in the direction of the slice proportionally to the
distance to its projected 1d optimal mapping. In \cite{paulin2020sliced}, the authors
have demonstrated the interest of such blue noise sampling in $[0,1)^d$ for Monte Carlo integration and Monte Carlo rendering.
This paper extends this approach to non-Euclidean metric spaces.

\paragraph*{Non-Euclidean Sliced Wasserstein Distance.} \dav{Bonet et al. 
extend the SW distance to Spherical  \cite{bonet2022spherical} and Hyperbolic metric spaces  \cite{bonet2022hyperbolic}}, by
replacing the Euclidean notions of lines and projections with the
Riemannian equivalent of projection over geodesics. Namely, the
spherical geodesics are great-circles of the sphere and geodesics
passing through the origin of any hyperbolic model are valid
replacements. 
{With these constructions at hand, authors perform various machine learning tasks where the SW distance is generally used as a data fitting loss or a meaningful metric to compare objects defined over such spaces.} 

\paragraph*{Blue Noise Mesh Sampling.} Blue noise sampling of 
surfaces in $\mathbb{R}^3$ is one of our targeted applications. 
On Euclidean domains, a classical approach to construct well-spread samples in a domain consists in
making sure that each pair of samples are separated  by at least a
given minimum distance. Dart throwing and its variations
 \cite{bridson2007fast} have naturally been extended to manifolds to
 achieve such Poisson disk sampling \cite{cline2009dart,Bowers2010,yuksel2015sample,guo2015efficient}.
 Alternatively,  Voronoi diagrams driven approaches  \cite{Liu:2009:OCV,Balzer2009CCPD} and 
 their restriction of discrete manifolds (triangular meshes in most cases), have been used to construct blue
 noise samples  \cite{liu2009centroidal,xu2012blue,Ahmed2016SimplePushPull,xin2016centroidal}. \dav{While focusing on remeshing applications, Peyré and Cohen \cite{peyre2006geodesic} have proposed an instrinsic sampling strategy that inserts samples one by one at the location maximizing the (geodesic) distance from the previous samples (approach denoted farthest-point, FP, below). While being efficient from an FMM approximation of the geodesic distance, this algorithm has a greedy approach and is not fit to sample generic non uniform densities.} 
 Starting
 from an initial sampling and pairwise (geodesic) distances between
 samples, \dav{Qin et al.}\cite{qin2017wasserstein} optimized samples position so that
 the regularized optimal transport distance between the samples and
 the uniform measure on the manifold is minimized. Particle based systems
 can be designed by optimizing the sample distribution on a mesh to unformize the distances between
 neighboring samples in ambient space, while staying close to the surface thanks
 to a projection operator \cite{tanaka2000sampling,zhong2013particle,jiang2015blue}. Samples could also be  optimized
 such that they capture the spectral content of the targeted surfaces  \cite{oztireli2010spectral}. In most cases,
 for efficiency purposes, the sampling is performed in ambient space and later projected onto the manifold.
{While those techniques can be very efficient in terms of blue noise quality when the mesh embedding to $\mathbb{R}^3$ is \emph{ambient-compatible} (no too-close sheets
 of meshes or large enough local shape diameter function \cite{shapira2008consistent},  Euclidean unit balls is a good approximation of the geodesic ones\ldots), we propose 
 an efficient purely intrinsic blue noise sampling that can deal with shapes with incorrect embedding (see Fig.~\ref{fig:teaser}).}

\section{Sliced optimal transport sampling on constant curvature manifolds}

{We first extend the SOTS approach defined on Euclidean domains,
  to the spherical and hyperbolic cases in arbitrary dimensions,
  respectively denote $\Sp^{d}$ and $\Hy^{d}$ (see Fig.~\ref{fig:models}).

To define the SOTS in such non-Euclidean spaces, we first need to
refine the notion of \emph{projection onto a straight line} \dav{as the projection of
a set of samples onto geodesic slices for the targeted model
(Sec.~\ref{sec:geodesic-slices}). Then we need to solve the matching
1d problem on the geodesic slice (Sec.~\ref{sec:sorting-along-slice}). These key ingredients are mostly
borrowed from \dav{Bonet et al. \cite{bonet2022spherical,bonet2022hyperbolic}} dedicated to the computation of $SW$ on $\Sp^d$ and $\Hy^d$. We extend these works with explicit formulas to perform the advection of the samples using group
action principle (Sec.~\ref{sec:trans-group-acti}) and Exp and Log maps
(Sec. \ref{sec:exp-log-map}).} Finally, Section
\ref{sec:riem-grad-desc} completes the algorithm describing the
extension of the gradient descent of the $SW_2$ energy. In Section
\ref{sec:geometric-median}, we describe a technical improvement of the
advection step on batches using a geometric median %
instead of an average as usually used in SOTS.}
We summarize the generic algorithm in Alg.~\ref{alg:nesots}. Note
that we consider a discrete target measure $\nu = \sum_{i=1}^m
\delta_{\yp_i}$ with a number of Diracs $m$ that may be greater than
$n$. This \dav{will}  be discussed in Section \ref{sec:non-unif-dens} to allow
the sampling of non-uniform densities.  \dav{Starting from line 5},
we \dav{thus} solve a balanced optimal transport problem as $\tilde{\nu}$ is a
random sampling of $\nu$ with exactly $n$ Diracs.

\begin{algorithm}\small
  \SetKwFor{ParallelFor}{parallel for}{do}{end}
\caption{Non Euclidean Sliced Optimal Transport Sampling -- NESOTS}\label{alg:nesots}
  \KwData{The  discrete target
  distribution $\nu = \sum_{i=1}^m \delta_{\yp_i}$, the number of
  iterations $K$, the batch size $L$, the gradient descent step $\gamma$}
  \KwResult{The discrete distribution $\mu^{(K)}$  
  after $K$ iterations.}
  $\mu^{(0)} = \text{SubSample}(\tilde{\nu},n)$ \tcp*{Init.}
\For{$ j \in [[1,K]]$}{
\ParallelFor(\tcp*[f]{Batch}){$ l \in [[1,L]]$}{
    $\tilde{\nu} = \text{SubSample}(\tilde{\nu},n)$ \tcp*{Sec.~\ref{sec:non-unif-dens}}
    $\theta$ = RandomSlice() \tcp*{Sec.~\ref{sec:geodesic-slices}}
    $\tilde{\nu}_\theta = P^\theta\left(\tilde{\nu}^l\right)$\tcp*{Sec.~\ref{sec:geodesic-slices}}
  $\mu_\theta = P^\theta\left(\mu^{(j)}\right)$\tcp*{Sec.~\ref{sec:geodesic-slices}}
  $T = \text{Solve1DOT}({\mu}_\theta$,$\tilde{\nu}_\theta)$ \tcp*{Sec. \ref{sec:sorting-along-slice}}
    \For{$ i \in [[1,n]]$}{
        $\gp = \Gamma_\theta\left(P^\theta\left(\xp^{(j)}_i\right),T\left(P^\theta\left(\xp^{(j)}_i\right)\right)\right)$ \tcp*{Sec.~\ref{sec:trans-group-acti}}
        $\dpp^l_i = \text{Log}_{\xp^{(j)}_i}\left (\gp\left ( \xp^{(j)}_i\right)\right)$ \tcp*{Sec.~\ref{sec:exp-log-map}}
    }   
}
    \ParallelFor{$ i \in [[1,n]]$}{
    $\dpp_i = \text{GeoMed}\left(\{\dpp^l_i\}_L\right)$ \tcp*{Sec.~\ref{sec:geometric-median}}
    $\xp_i^{(j+1)} = \text{Exp}_{\xp^{(j)}_i}\left(\gamma\, \dpp_i\right)$ \tcp*{Sec.~\ref{sec:riem-grad-desc}}
    }   
}
\Return $\mu^{(K)}=\sum_{i=1}^m \delta_{\xp^{(K)}_i}$
\end{algorithm}

\subsection{Geodesic slices and projections}
\label{sec:geodesic-slices}
The first step is to find an equivalent to straight lines in the Euclidean space. The most natural choice is a geodesic passing
through the origin of the model. In both $\Sp^d$ and $\Hy^d$ cases, such an
object can be obtained by the intersection of a plane with the
canonical embedding of each space in $\mathbb{R}^{d+1}$.
\paragraph*{Spherical geometry.} As proposed by Bonet et
al. \cite{bonet2022spherical}, random slices are defined by the
intersection of $\Sp^d$ by uniformly
sampled Euclidean 2D planes in $\mathbb{R}^{d+1}$ passing through
the origin. This is done  by generating two $(d+1)-$dimensional
 vectors with components in $\mathcal{N}(0,1)$, that we  orthonormalize (by Gram-Schmidt or
Givens rotations). We denote by $\theta = \{\ep_1,\ep_2\}$ the two
vectors in $\mathbb{R}^{d+1}$ generated by this process. Such basis of the plane allows defining the projection in $\mathbb{R}^{d+1}$ onto the associated
subspace $\Span(\ep_1,\ep_2)$:
    \begin{equation}
        \Pi^\theta(\xp) = \langle \xp , \ep_1 \rangle \ep_1 + \langle \xp , \ep_2 \rangle \ep_2\,.
    \end{equation}
The  projection onto the great circle $= \Span(\ep_1,\ep_2)\cap \Sp^d$ becomes
\begin{equation}
    P^\theta(\xp) := \normalize{\Pi^\theta(\xp)}\,.
\end{equation}
\paragraph*{Hyperbolic geometry.}

The $d-$dimensional hyperbolic plane $\Hy^d$ admits many isometric
models (\emph{e.g.} the Poincaré disk or the Lorentz's hyperboloid models) \cite{lee2006riemannian}. For the sake of
simplicity of the associated formulas and numerical reasons, we will be using the
hyperboloid model, i.e., the upper sheet of the hyperbola
$$\Hy^d :=\{\xp\in\Euc^{d+1}, \dotL{\xp}{\xp} = -1\}\,,$$
where $\dotL{\xp}{\yp} :=
\sum_{i=1}^{d} x_iy_i - x_{d+1}y_{d+1}$ {is the Lorentzian dot product}. We denote by $\xO$ the origin
of the hyperbola (red dot in Fig.~\ref{fig:models}), i.e., $\xO = \begin{pmatrix} 0, \dots, 0, 1\end{pmatrix}^t$.
We follow \dav{Bonet et al.} \cite{bonet2022hyperbolic} by defining the projection on the
geodesic obtained as the intersection between a 2D plane containing $\xO$ and the
hyperboloid.
The sampling of uniform slices is achieved by sampling uniformly the space
 orthogonal to $\xO$, i.e. $\dpp \sim \mathcal{U}(\Sp^{d}\times\{0\})$.
We then have the projector 
\begin{equation}
    P^{\theta}(\xp) := \frac{\Pi^{\theta}(\xp)}{\sqrt{-\dotL{\Pi^\theta(\xp)}{\Pi^\theta(\xp)}}}\,,
\end{equation}
where we denote by $\theta := \{\dpp,\xO\}$ the generator of the 2D slice
in $\Hy^d$.

\subsection{Solving the discrete 1D Wasserstein problem}
\label{sec:sorting-along-slice}

As we will need to evaluate the transport cost on projected samples onto the sliced $\theta$, 
we need to clarify the distances between two points in $\Sp^d$ or $\Hy^d$, and the coordinate on their 
projection onto $\theta$, denoted $t_\theta(\xp)$,  the signed geodesic distance to {a given} origin in $\theta$ .

\paragraph*{Spherical geometry} On the $d-$dimensional unit sphere, geodesics are great
circles (intersection of a 2-plane passing through the origin, and $\Sp^d$).  The geodesic distance between two points 
$\xp,\yp\in\Sp^d$
is simply  the angle
between the two vectors from the origin to the points
\begin{equation}
    d_\Sp(\xp,\yp) := \acos(\dotE{\xp}{\yp})\,.
\end{equation}
As projections lie on a circle, any origin on $\theta$ can be considered to define $t_\theta$. If $\thetap = \{\ep_1,\ep_2\}$, we use
\begin{equation}
    t_\theta(\xp) := \frac{\pi + \atan(\dotE{\ep_2}{\xp},\dotE{\ep_1}{\xp})}{2\pi}\,.
\end{equation}

 On $\Sp^d$, the optimal transport problem needs to take into account the periodicity of the space, and its associated coordinate systems. Fortunately, it can be shown \cite{delon2009transportation} that the problem still boils down to a simple sorting of the samples coordinates $t_\theta$ provided that the circle is identified to the Real line through an optimal cut. Finding the optimal cut can be formulated as a weighted median problem, as detailed in \dav{Cabrelli et al.} \cite{cabrelli1998linear}, and admits a
$\mathcal{O}(n\log(n))$ solution. For some $\mu$, $\nu$ in $\Sp^d$ and $\xp\in\mu$, the map $T(P^\theta(\xp))$ denotes the 
optimal assignment on the slice $\theta$ of $\xp$ to some $\yp\in\nu$.

\paragraph*{Hyperbolic geometry} On $\Hy^{d}$, the geodesic distance between two points is 
\begin{equation}
    d_\Hy(\xp,\yp) := \acosh(-\dotL{\xp}{\yp})\,.
\end{equation}
Since the slice is directed by $\dpp$, we define the geodesic distance coordinate induced by $\dpp$
\begin{equation}
  t_\thetap(\xp) := \sgn(\dotE{\xp}{\dpp})d_\Hy(\xO,\xp)\,.
\end{equation}
On $\Hy^d$, the optimal assignment is simply obtained by sorting the projected samples
 on $\theta$ and mapping the first projected sample in $P^\theta_\#\mu$ to the first one in  $P^\theta_\#\nu$ (with respect to $t_\theta$), similarly 
 to the Euclidean case.

\subsection{Transitivity and group action}
\label{sec:trans-group-acti}

In  the Euclidean space, samples are advected  by a simple translation in the straight line
direction by the distance $t_\theta(\xp) - t_\theta(T(P^\theta(\xp)))$. 
In spherical  \dav{(Eq.~(\ref{eq:gamma1}))} and hyperbolic  \dav{(Eq.~(\ref{eq:gamma2}))} domains, we rely on group
actions. More precisely, we are interested in group actions that
preserve the geodesics.
\paragraph*{Spherical Geometry} The right group to act on the sphere
is $SO(d)$, i.e., the group of all $d-$dimensional rotations. One can build
the rotation that maps a point $\xp$ to a point $\yp$ in $\Sp^d$ simply by building
the 2D rotation in their common span, $\Span(\{\xp,\yp\})$, i.e
\begin{align*}
    \begin{pmatrix}
    \cos(\varphi) & - \sin(\varphi) \\ \sin(\varphi) & \cos(\varphi)
    \end{pmatrix}\,,
\end{align*}
for some $\phi\in\mathbb{R}$. To make sure that the part of the vector
orthogonal to span$(\xp,\yp)$ is left unchanged and to avoid building
the $d\times d$ matrix, we decompose any vector $\wpp$ in the
orthonormal basis given as the result of the Gram-Schmidt algorithm
applied to $\xp$ and $\yp$. Leading to
\begin{align}
    \Gamma_\theta(\xp,\yp): \wpp \rightarrow \wpp^{\perp} &+ \xp(\cos(\varphi)w_x - \sin(\varphi)w_y) \nonumber\\&+ \tilde{\yp}(\sin(\varphi)w_x + \cos(\varphi)w_y)\,,\label{eq:gamma1}
\end{align}
where $\tilde{\yp} = \yp - \dotE{\xp}{\yp}\xp$, $w_x =
\dotE{\wpp}{\xp}$, $w_y = \dotE{\wpp}{\tilde{\yp}}$, $\wpp^\perp$ is
the component of $\wpp$ orthogonal to $\Span(\{e_1,e_2\})$ and $\varphi =
d_\Sp(\xp,\yp)$. One can verify that we have $
\Gamma_\theta(\xp,\yp)(\xp) = \yp$.
It is also possible to check that a rotation of $\varphi$ degree along the slice
$\theta$ applied to $\xp$ will offset $t_\theta(\xp)$ by $\varphi$
(modulo $1$). Hence, it is indeed a translation along the slice, which
is the behavior we wanted to translate from the Euclidean setting.

\paragraph*{Hyperbolic Geometry} 

As a direct analogy, translations along hyperbolic slices are hyperbolic rotations, i.e., the elements of the Lorentz group $SO_0(d-1,1)$ (standard rotations preserve the Euclidean scalar product whereas hyperbolic ones preserve $\dotL{\cdot}{\cdot}$, hence the hyperboloid).
Computationally, it is very similar to
the spherical case, we want to apply the following 2D rotation in the $\Span(\xp,\yp)$: 
\begin{align*}
    \begin{pmatrix}
    \cosh(\varphi) & \sinh(\varphi) \\ \sinh(\varphi) & \cosh(\varphi)
    \end{pmatrix}\,,
\end{align*}
leading to the analogous decomposition along the right subspaces:
\begin{align}
    \Gamma_\theta(\xp,\yp): \wpp \rightarrow \wpp^{\perp} &+ \dpp(\cosh(\varphi)w_d  + \sinh(\varphi)w_0) \nonumber\\&+ \xO(\sinh(\varphi)w_d + \cosh(\varphi)w_0)\,,\label{eq:gamma2}
\end{align}
where $\dpp = \normalize{\Pi_{x_0^\perp}(\yp-\xp)}$, $w_d =
\dotE{\wpp}{\dpp}$, $w_{0} = \dotE{\wpp}{\xO}$, $\wpp^\perp$ is the
component of $\wpp$ orthogonal to $\Span(\xO,\dpp)$ and $\varphi =
d_\Hy(\xp,\yp)$. The only difference being that we decompose along
$\xO$ and $\yp-\xp$ instead of directly $\xp$ and $\yp$ (which gives
the same span) to make sure that the points remain on the
hyperboloid. We also have $\Gamma_\theta(\xp,\yp)(\xp) = \yp$.

\subsection{Exp and Log Maps}
\label{sec:exp-log-map}

\begin{figure}
  \centering
  \begin{overpic}[width=8cm,trim=0 14cm 0 0,clip]{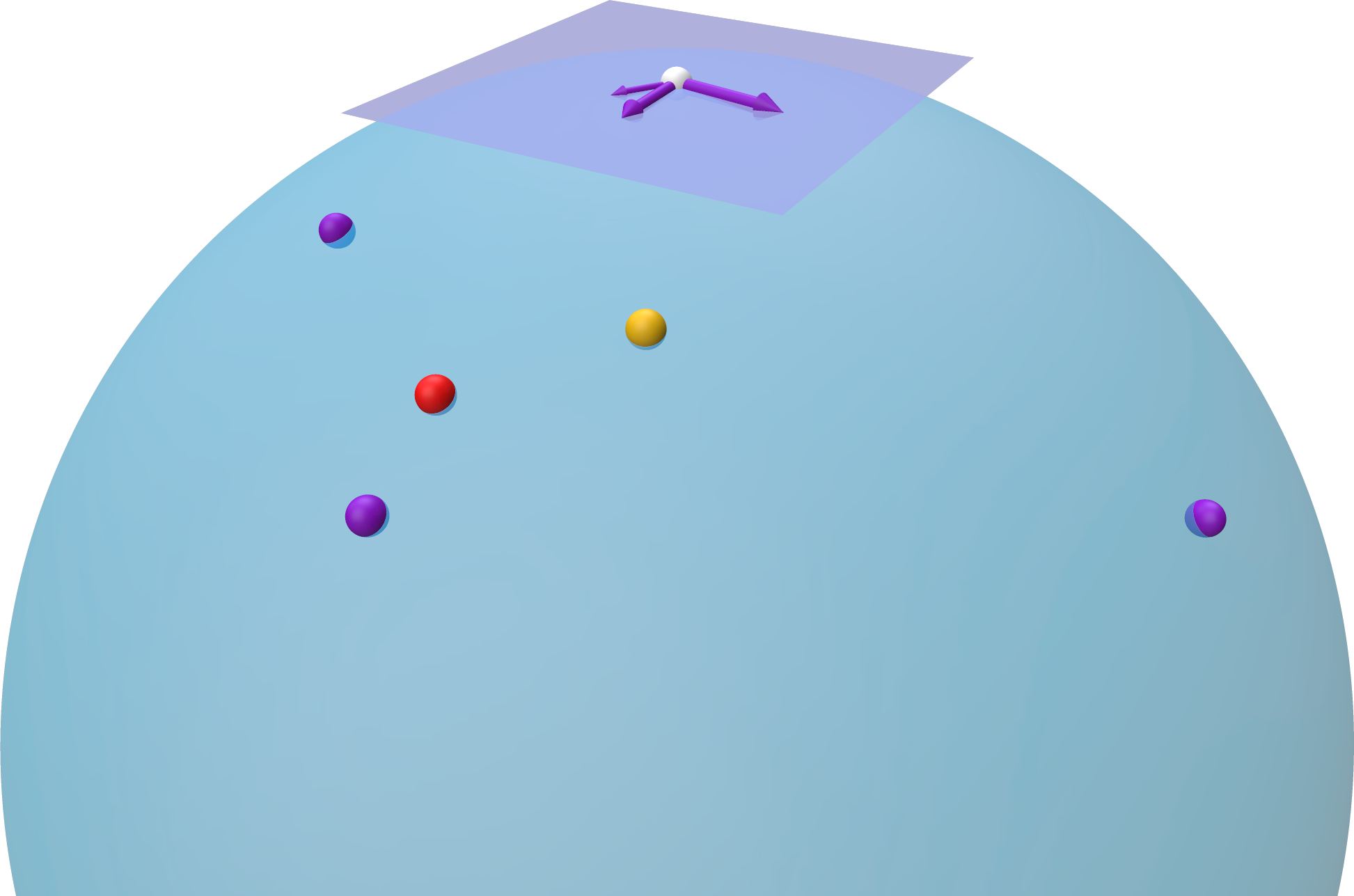}
    \put(74,12){$\yp= \Exp_\xp(\vp)$}
    \put(48,42){$\xp$}
    \put(32,37){\color{gray}$TM_\xp$}
    \put(59,35){$\vp=\Log_\xp(\yp)$}
  \end{overpic}
  \caption{\textbf{Exp and Log maps:} on $\Sp^2$, \batou{ the orange point
    is the point obtained by iteratively going in the average of the Logs $x_{n+1} = \Exp_{x_n}(\frac{\gamma}{n} \sum_i \Log_{x_n}(y_i))$, which is equivalent to Fréchet means, 
    whereas the red one is obtained by going in the geometric median of the directions $x_{n+1} = \Exp_{x_n}(\gamma\,\,\text{GeoMed}(\{\Log_{x_n}(y_i)\}_i))$.\label{fig:logexp}}}
\end{figure}

Beside group actions, Exp and Log maps are key ingredients in
  Riemannian geometry \cite{lee2006riemannian} (see illustration Fig.~\ref{fig:logexp}).
The $\Exp_\xp(\vp)$ map allows one to follow the geodesic $\gamma$,
satisfying $\gamma(0) = \xp$ and $\dot{\gamma}(0) = \vp \in TM_\xp$, i.e.,
following the most natural path going from $\xp$ with initial direction
and velocity $\vp$ from $t=0$ to $t=1$.
Conversely,  the $\Log_{\xp}(\yp) \in TM_\xp$ map, the inverse of
$\Exp_\xp$, gives the direction (and velocity) 
to go from $\xp$ to  $\yp$, i.e. $\Exp_\xp(\Log_\xp(\yp)) = \yp$.
In $\Sp^d$ and $\Hy^d$, Exp and Log maps admit  closed form expressions.
\paragraph*{Spherical geometry.} If $\Pi_{TM_\xp}$ denotes the projections from $\mathbb{R}^d$ onto the
tangent space of $\Sp^d$ at $\vp$, we have
\begin{align} 
    \Exp_\xp(\vp) &= \cos\left(\|\vp\|)\xp + \sin(\|\vp\|\right)\normalize{\vp}\,, \\
    \Log_\xp(\yp) &= \normalize{\Pi_{TM_\xp}(\yp-\xp)}d(\xp,\yp)\,,
\end{align}
(see \dav{Alimisis et al.'s supplemental \cite{alimisis2021distributed}}).

\paragraph*{Hyperbolic geometry.}
In the Lorentz hyperbolic model, we have similar expressions (see {\em e.g.} \dav{Dai et al.} \cite{dai2021hyperbolic}):
\begin{align} 
    \Exp_\xp(\vp) &= \cosh(||\vp||_\mathbb{L})\xp + \sinh(||\vp||_\mathbb{L})\frac{\vp}{||\vp||_\mathbb{L}}\,, \\
    \Log_\xp(\yp) &= \frac{\acosh(-\dotL{\xp}{\yp})}{\sqrt{\dotL{\xp}{\yp}^2-1}}(\yp + \dotL{\xp}{\yp}\xp).
\end{align}

\subsection{Stochastic Riemannian gradient descent}
\label{sec:riem-grad-desc}

\begin{figure}
  \centering 
   \begin{tabular}{ccc}
   {\includegraphics[width=2.2cm]{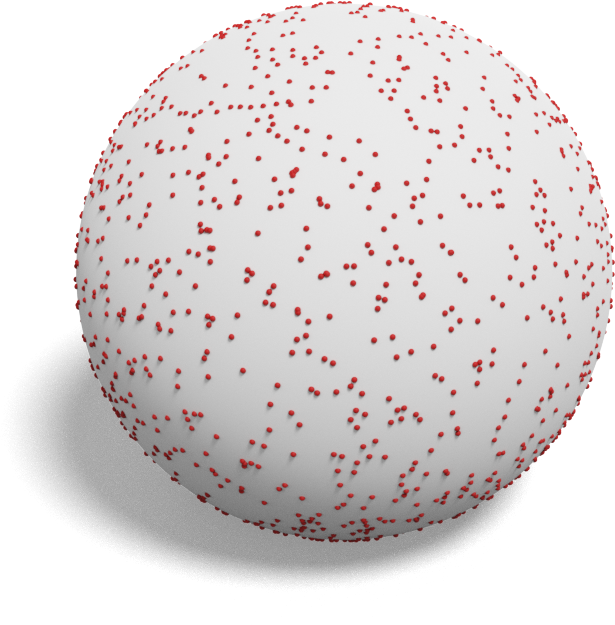}}&
   {\includegraphics[width=2.2cm]{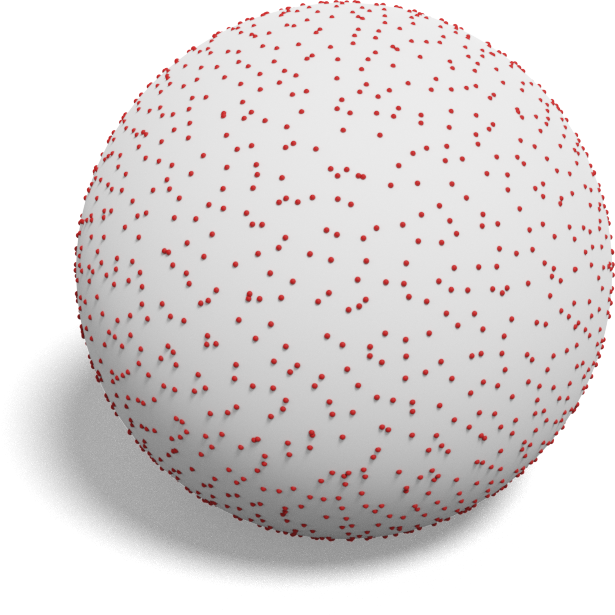}}&
   {\includegraphics[width=2.2cm]{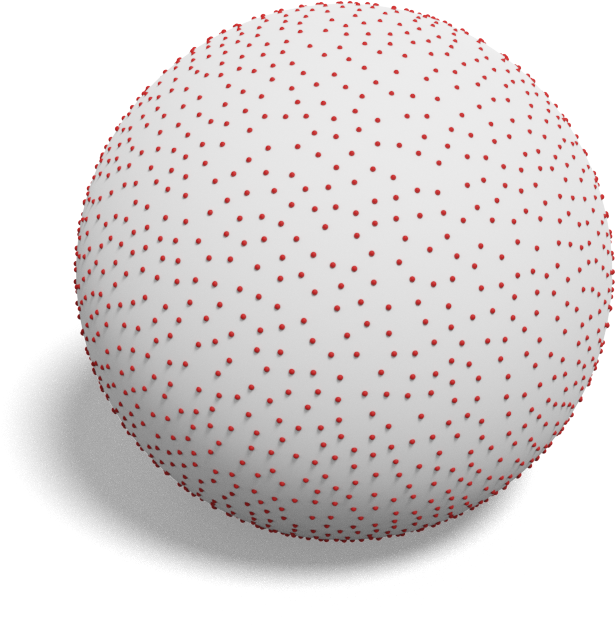}}\\
   WN& Stratified&Poisson disk\\
   {\includegraphics[width=2.2cm]{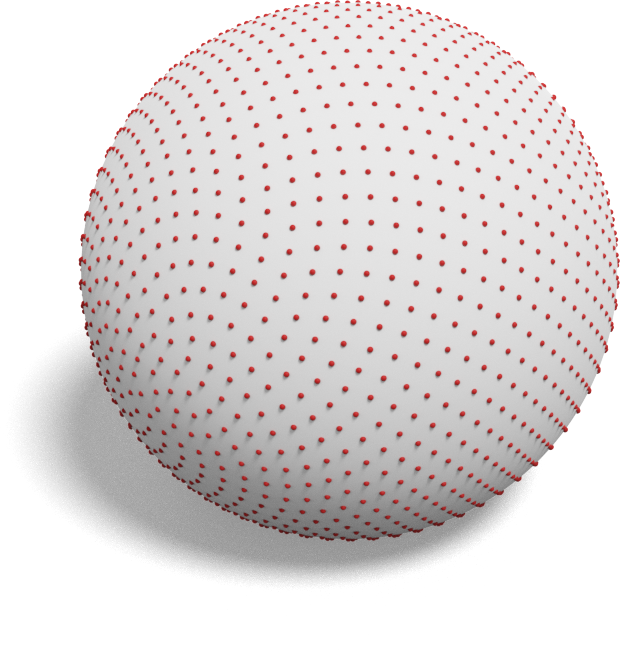}}&
   {\includegraphics[width=2.2cm]{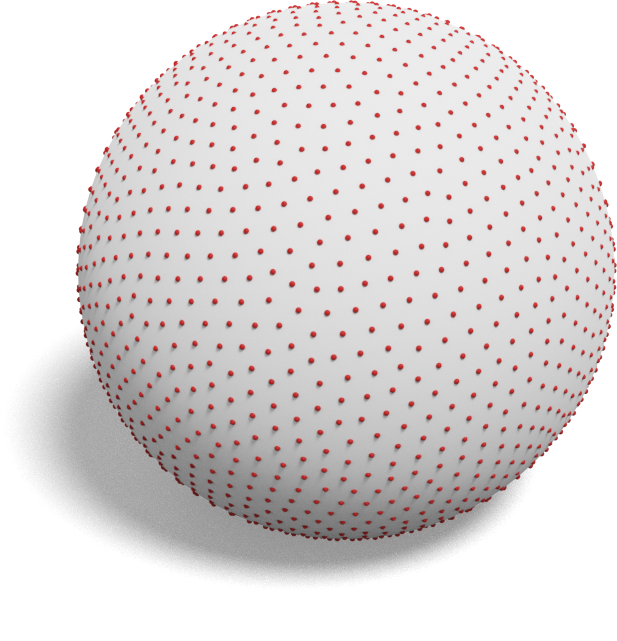}}&
   {\includegraphics[width=2.2cm]{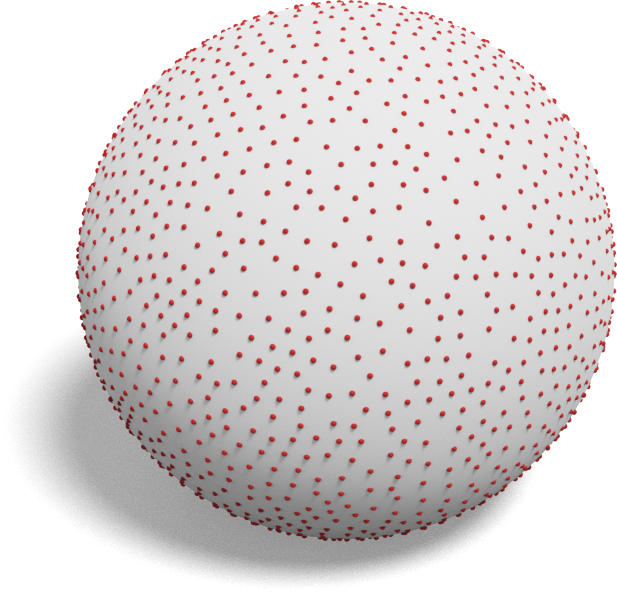}}\\
   Spherical Fibo.&CVT& FP 
   \end{tabular}\\
   \begin{tabular}{cc}\raisebox{.75cm}{\begin{overpic}[width=2.2cm]{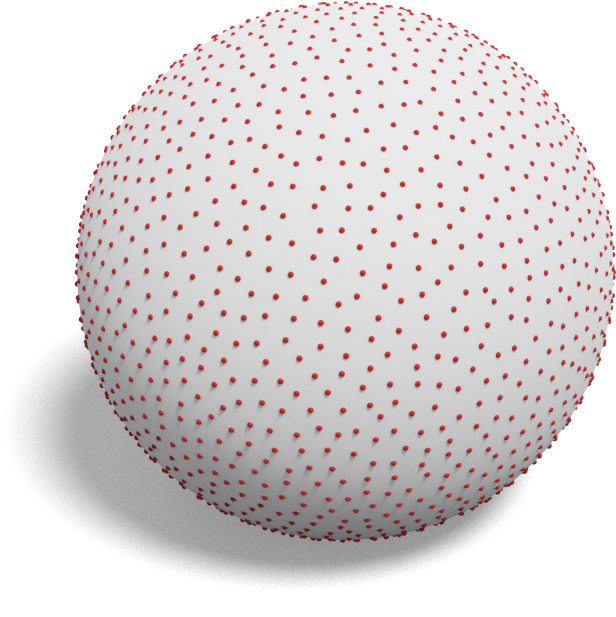}
   \put(30,-10){NESOTS}\end{overpic}}&{\includegraphics[width=6cm]{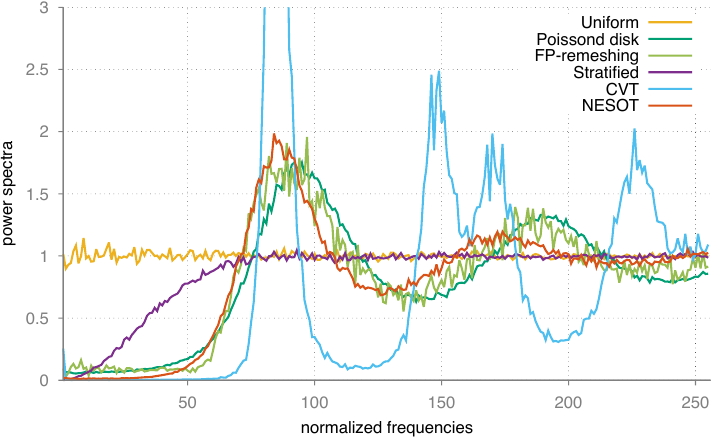}}
   \end{tabular}
\caption{\textbf{Blue noise on the sphere.} On $\Sp^2$, we evaluate the blue noise property of our
  sampling (2048 samples). Our result as to be compared to a uniform
  sampling, a stratified sampling
  using a healpix spherical structure \cite{Pilleboue:2015:VAMCI}, a Poisson disk sampling, a spherical Fibonacci
  sequence \cite{keinert2015spherical}, and a Lloyd's relaxation approach
  (Centroidal Voronoi Tesselation\dav{, CVT}) \cite{liu2009centroidal}, \dav{and a geodesic farthest point greedy strategy \cite{peyre2006geodesic} (FP)} . The graph
  corresponds to the angular power spectra of the spherical
  harmonic transform of the point sets (except for spherical
  Fibonacci whose regular patterns make the spectral analysis less relevant) . As discussed in \dav{Pilleboue et al.}
  \cite{Pilleboue:2015:VAMCI}, our sampler exhibits correct blue noise
  property with low energy for low frequencies, a peak at the average
  distance between samples and a plateau with few oscillations for
  higher frequencies.\label{fig:spectra}}
 \end{figure}

\batou{
In Euclidean SOTS, when optimizing point sets for blue noise sampling, one can compute a descent direction of the SW energy for each point by
averaging each advection computed for a given number of slices (batch
size $L$ in Alg.~\ref{alg:nesots}), hence
recovering a {mini-batch} stochastic gradient descent. On non-Euclidean domains,
the advected positions cannot be simply averaged. We propose to use a stochastic Riemannian gradient descent (SRGD) approach combining the gradients obtained in each batch in the tangent plane of each sample \cite{boumal2023introduction}. In standard SRGD this would be done by taking the average of the gradients 
\begin{equation}
  \dpp_i := \frac{1}{L}\sum_{l=1}^L \dpp_i^l \,,
\end{equation}
but we instead use the geometric median, see \ref{sec:geometric-median}.
In our case, $\dpp_i^l := \Log_{\xp^{(j)}_i}\left (\gp(\xp^{(j)}_i)\right)$, where,
 following the notations of Alg.~\ref{alg:nesots}, $\gp$ is the map that advects
 the point $\xp_i^{(j)}$ in the $\theta$ \emph{direction} following the 1D assignment obtained from the projection onto $\theta$. 
Once the descent direction is computed for each sample, we advect the points using the
$\Exp$ map by an, exponentially decaying, step size $\gamma$:
\begin{equation}
    \xp^{(j+1)}_i = \Exp_{\xp^{(j)}_i}\left(\gamma\,\,\dpp_i\right)\,.
\end{equation}
}
Note that in the Euclidean setting, this boils down to the original SOTS algorithms
\cite{bonneel2015sliced} for blue noise sampling in $[0,1)^d$. {As a first experiment, Figure \ref{fig:spectra} compares the blue noise characteristics of the 
uniform sampling of using NESOTS and classical point patterns on $\Sp^2$ \cite{Pilleboue:2015:VAMCI}.}

\subsection{Non-uniform densities}
\label{sec:non-unif-dens}

\begin{figure}
  \begin{center}
\subfigure[]{\includegraphics[width=3.5cm]{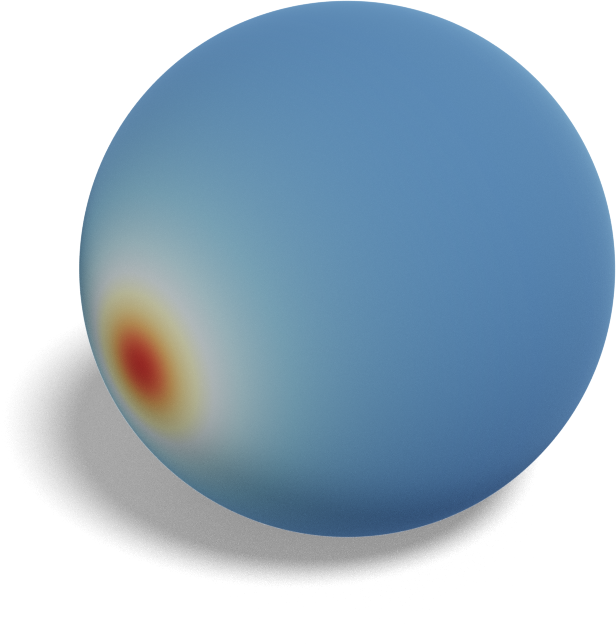}}
\subfigure[]{\includegraphics[width=3.5cm]{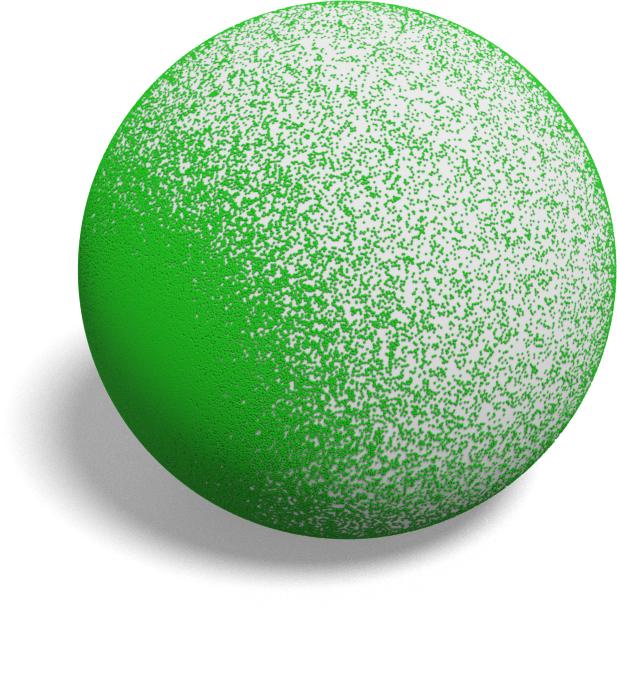}}\\
\subfigure[]{\includegraphics[width=3.5cm]{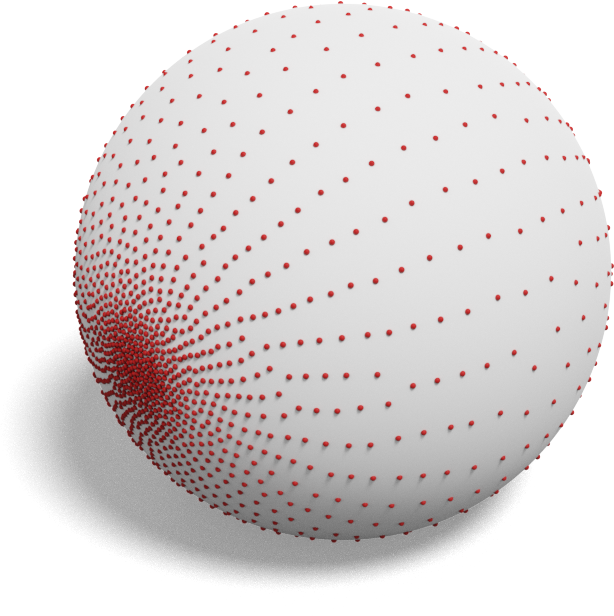}}
\subfigure[]{\includegraphics[width=3.5cm]{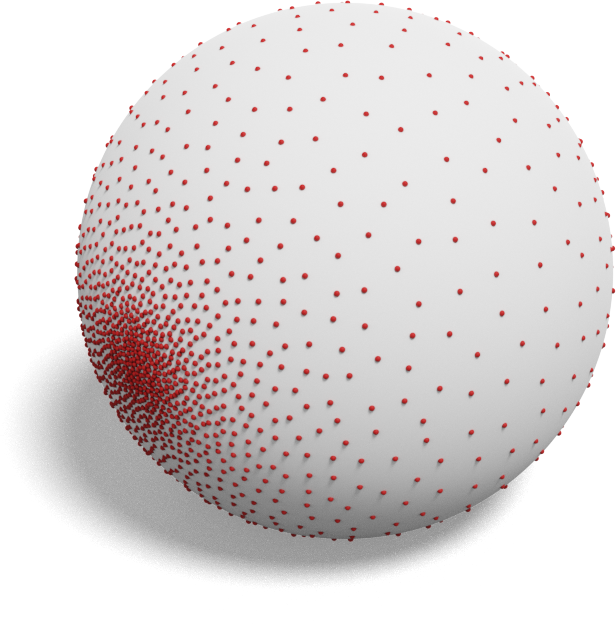}}
  \end{center}
\caption{\textbf{Non-uniform measure sampling:} given a non-uniform probability measure $\phi$ in  $\Sp^2$ $(a)$, we
  first construct a discrete  measure $\nu\sim\phi$ with a large
  number of samples, 2048 samples here $(b)$.  Figures $(c)$ and $(d)$
  are the output of the NESOTS algorithm for 2048 samples ($L=32$, $K=300$),  when
  averaging the directions during the advection $(c)$, or using the
  geometric median $(d)$. While both distributions approximate the
  density, the latter provides a more stable result without sample
  alignment artifacts.\label{fig:nonunif}}
  \end{figure}

When dealing with continuous non-uniform measures $\phi$ using a
sliced approach (\emph{e.g.} importance sampling Monte Carlo rendering, image stippling), we would first
need to  have a closed-form formulation of the Radon transform of the
target  measure of $\phi$ along the slice $\theta$, as discussed \dav{Paulin et al.}
\cite{paulin2020sliced} for the uniform measure in $[0,1)^d$. To overcome such issue, Sala{\"u}n et al. \cite{salaun2022scalable} have used a  binning strategy of the target points across $n$ adaptive bins that follow the target distribution.  We
further simplify  this approach on $\Sp^d$ and $\Hy^d$ using an empirical
  approximation of  $\phi$ from a discrete measure $\nu$ with a large
  number of samples $m$ (see Fig.~\ref{fig:nonunif}). The key idea of Alg.~\ref{alg:nesots} is to
  start from $\nu$ with $m\gg n$, and to uniformly pick $n$ samples from $\nu$
  at each slice (line 5). As long as $\nu \sim \phi$,
   this does not
  affect the minimization of the SW energy, while allowing a lot of
  flexibility with respect to the applications (see below) and keeping
  a balanced $n$-to-$n$ 1d optimal transport problem to solve.

 \subsection{Geometric median}
\label{sec:geometric-median}

In our experiments, we observe that when targeting non-uniform measures, artifacts may appear during the 
gradient descent (\emph{e.g.} alignment of samples as illustrated in Fig.~\ref{fig:nonunif}-$c$).
Some approaches handle this fact with a
more robust advection computation, such as \dav{Sala{\"u}n et al.} \cite{salaun2022scalable}
but they all require a non-negligible computational overhead,
proportional to the input size (for example taking $m =
kn$). To overcome this problem without adding limited extra computations, instead of
taking the mean of the descent directions, we compute their geometric
median. The idea arose from the analogy between the arbitrary bad
batches that occurs with poor quality subsamples $\tilde{\nu}^l$ and
malignant voters in voting systems, see
\cite{el2023strategyproofness}. The geometric median can be computed
very efficiently, in practice using the Weiszfeld algorithm \cite{weiszfeld1937point}, see Appendix \ref{sec:weiszfelds-algorithm}.

\subsection{Real projective plane sampling}
\label{sec:RPPS}

A slight modification of the NESOTS algorithm on the sphere allows sampling any density defined on the real projective plane $\mathbb{P}^d$ in the same blue noise way. Such sampling might have great use in graphics applications since many geometric objects are defined up to signs (such as directing vectors of lines or plane normals). Applications are detailed in section 6.

\section{Intrinsic discrete manifold sampling}
\label{sec:intr-surf-sampl}

\definecolor{babyblueeyes}{rgb}{0.63, 0.79, 0.95}

\begin{figure*}\centering
 
  \begin{overpic}[width=14cm]{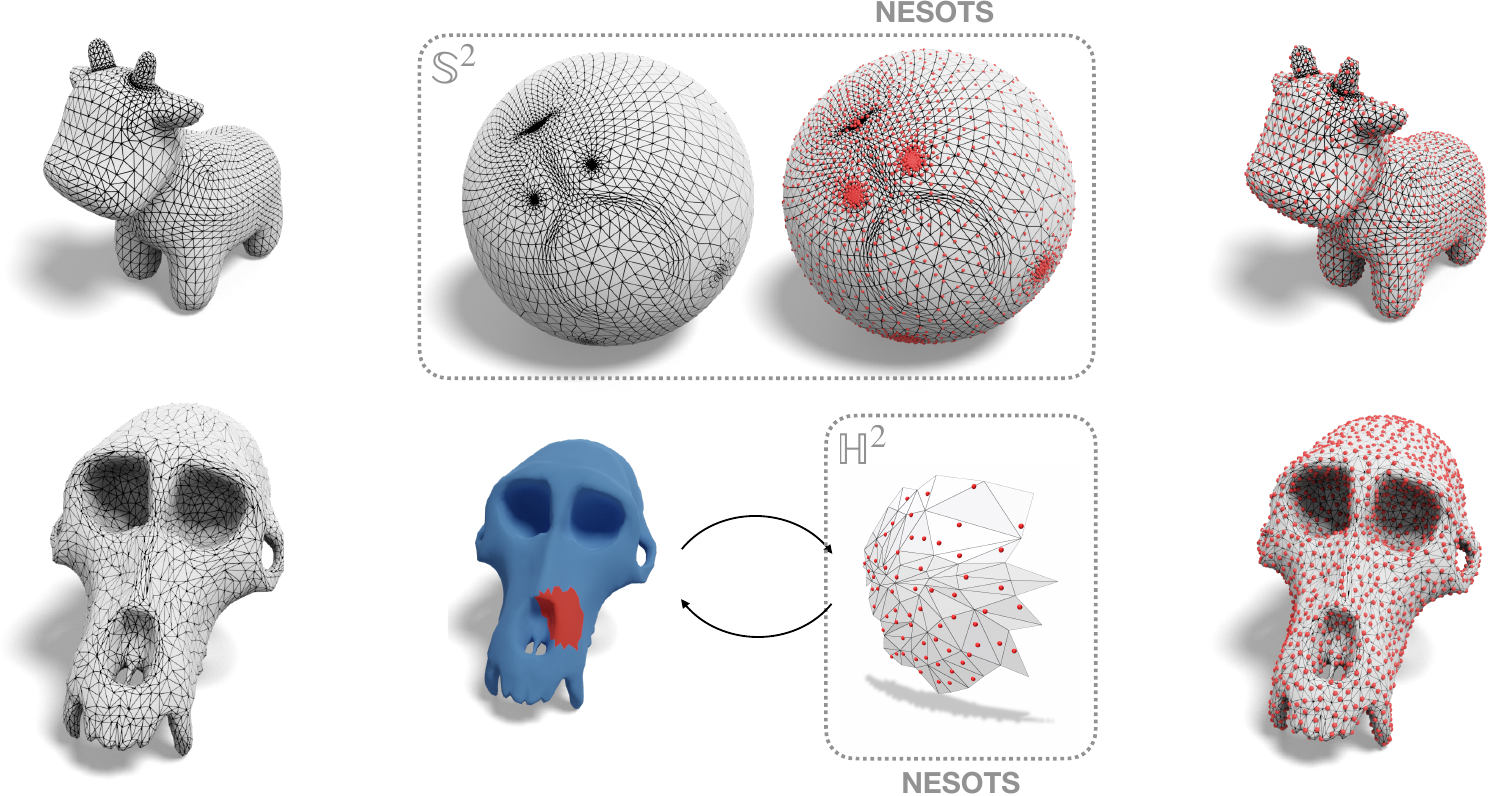}
    \put(21,40){\Large$\overset{\psi}{\longrightarrow}$}
    \put(77,40){\Large$\overset{\psi^{-1}}{\longrightarrow}$}
    \put(49,20){$\psi_i$}
    \put(49,7){$\psi_i^{-1}$}
  \end{overpic}
\caption{\textbf{Overall pipeline of our intrinsic discrete manifold
  sampling}: starting from an input shape, we conformally embed the discrete structure  onto either
  $\Sp^2$ for 0-genus surfaces, or local patches to $\Hy^2$ for higher genus one. Then, the NESOTS (Alg.~\ref{alg:nesots}) is used (globally or locally) to
  blue noise sample the embedded structure targeting a measure taking
  into account the metric distortion.\label{fig:pipeline}}
\end{figure*}

As a first application, we demonstrate the interest of the
non-Euclidean sliced optimal transport approach for intrinsic sampling
of meshes in $\mathbb{R}^3$. Given a (closed) mesh $\mesh$, the core idea is to
construct an injective map $\psi$ from $\mesh$ to $\Sp^2$ or  $\Hy^2$,
to apply NESOTS on these domains to sample the image of the uniform 
measure $\mathcal{U}(\mesh)$ on the mesh by $\psi$ and to pull back the samples onto
$\mesh$ with $\psi^{-1}$.
{Fig.~\ref{fig:pipeline} gives an illustration of this general pipeline.}

For surfaces in $\mathbb{R}^{d}$, $\psi$ can be built as a conformal map
through the uniformization theorem \cite{abikoff1981uniformization}. {For short, any Riemannian 
surface of genus $g$ admits a constant Gaussian curvature metric: spherical metric if $g=0$ ($\Sp^{d-1}$, positive constant curvature space), a flat metric  f $g=1$ ($\mathbb{R}^{d-1}$, zero-curvature space) and an hyperbolic metric for $g\geq 2$ ($\Hy^{d-1}$, negative curvature space).}
In the discrete setting, $\mesh$ and $\mesh'$ are
discrete conformal equivalent if the edge lengths $l_{ij}$ and
$l'_{ij}$ are such that
  $l'_{ij} = \exp^{(u_i+u_j)/2}l_{ij}\,,$
for some conformal factors $\{u_i\}\in\mathbb{R}$ on vertices \cite{springborn2008conformal,bobenko2015discrete,gu2018discrete,schmidt2020inter}. {In the following, we specifically target the $g=0$ and $g\geq 2$ cases.}

\begin{figure}[!h]\centering
  {\includegraphics[width=2.6cm]{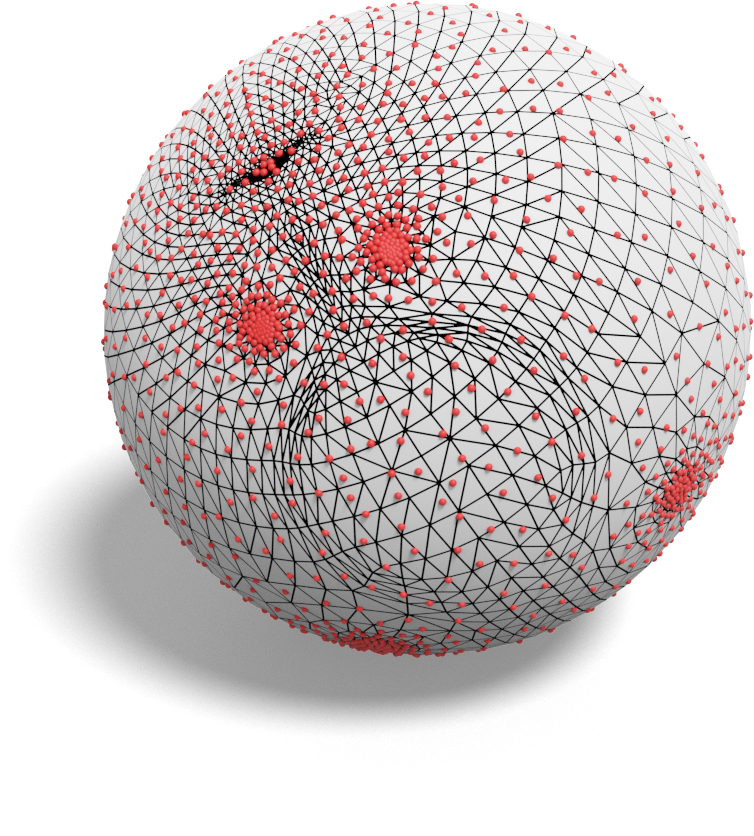}}
{\raisebox{0.5cm}{\includegraphics[width=2.6cm]{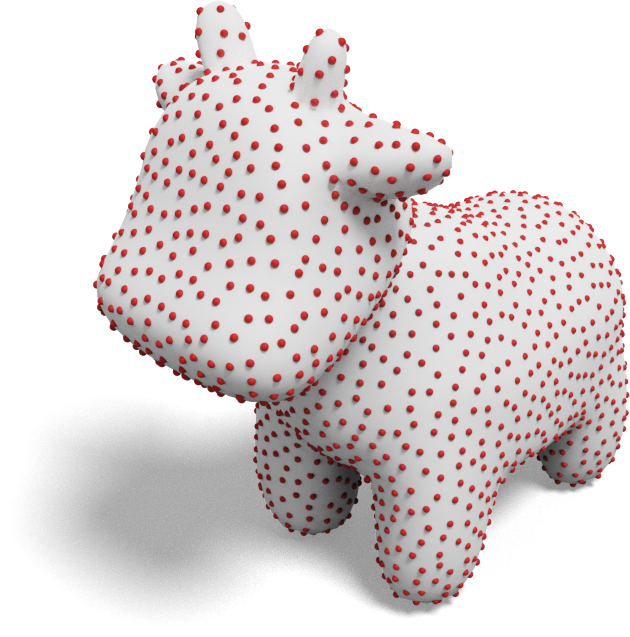}}}\\
  {\includegraphics[width=2.6cm]{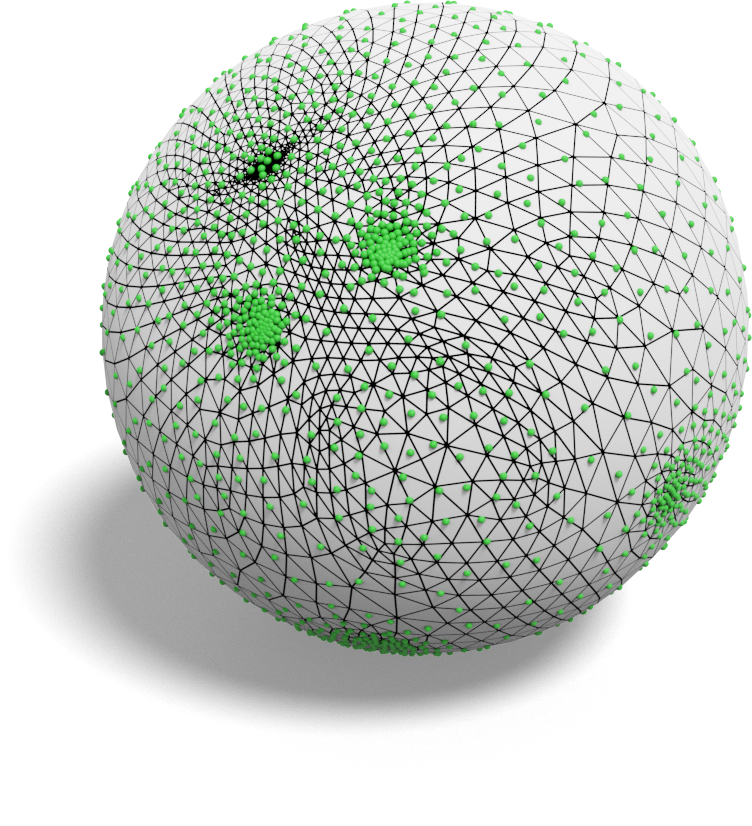}}
  {\raisebox{0.5cm}{\includegraphics[width=2.6cm]{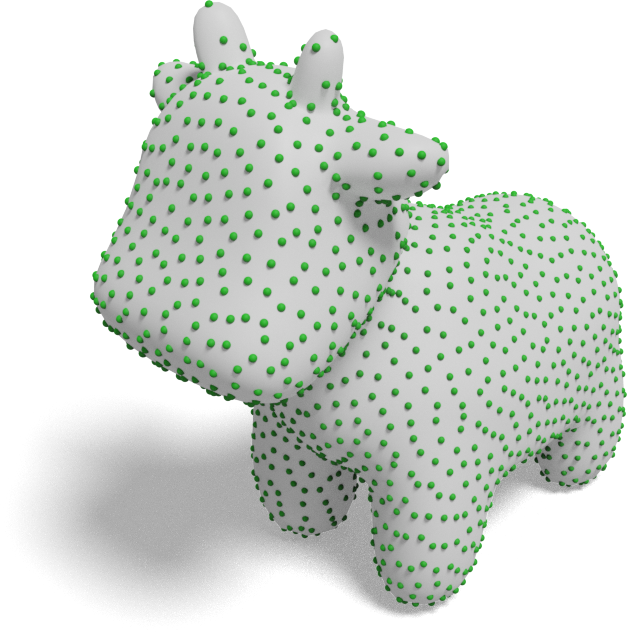}}}
  \caption{\dav{\textbf{Sampling using a non-conformal spherical mapping:} first, we recall the NESOTS samplings using CEPS conformal maps (first row). In green, we have updated the mapping using some Laplacian smoothing steps on the sphere, resulting comparable sampling (second row).\label{fig:spotmappings}}}
\end{figure}

\dav{
Note that in our pipeline, we do not explicitly require the map to be
conformal. Any injective map between the mesh and the target
space could be considered. 
We focus here on conformal maps as theoretical guarantees of existence and efficient algorithms to compute them exist. In Fig.~\ref{fig:spotmappings}, we illustrate that comparable blue noise sampling can be obtained non-conformal maps.

In the next section, we describe the sampling algorithm on the sphere,
also illustrated in Fig.~\ref{fig:pipeline}.
Section \ref{sec:local-hyperb-embedd}
focuses on high genus surfaces using an
iterated local hyperbolic embedding.
}
Our samples minimize the sliced transport energy to the target measure with respect to the ground metric of the embedded space ($\Sp^d$ or $\Hy^d$), not the intrinsic metric of $\mesh$. Yet, from the regularity of the conformal maps we observe that blue noise characteristics are preserved when pulled back from the embedded space to $\mesh$ (see Sec.~\ref{sec:implem}).

\subsection{Global spherical embedding}
\label{sec:sphere_sampl-alg}

The construction of the mapping $\psi$ through the uniformization
theorem depends on the genus $g$ of $\mesh$. For the sake of simplicity,
we start with the spherical case i.e., $g=0$. By the 
uniformization theorem, a conformal map exists from $\mesh$ to
$\Sp^2$. Here, we take advantage of the robust tools provided by
Gillespie et al. \cite{Gillespie:2021:DCE} to construct a bijective  conformal 
map $\psi:\mesh\to\Sp^2$, allowing a global optimization.

\begin{algorithm}
  \caption{Intrinsic Spherical blue noise surface sampling}
  \label{alg:ibnss}
  \KwData{$\mesh$, $\nu$, $m$, $n$, $K$, $L$ and $\gamma$ (see Alg.~\ref{alg:nesots})}
  $\mesh_G = \text{BuildMapping}(\mesh,\Sp^2)$ \;
  ${\nu}_G = \text{sampleMeshFaces}(\mesh_G,\nu,m)$ \;
  $\tilde{\mu_G} = \text{SubSample}(\nu_G,n)$ \;
  $\mu_G = \text{NESOTS}(\tilde{\mu_G},{\nu}_G,K,L,\gamma)$ \tcp*{Alg.~\ref{alg:nesots}}
  $\mu = $ MapToMesh$(\mu_G,\mesh,\mesh_G)$ \tcp*{Alg.~\ref{alg:mmbm}}
  \Return{$\mu$}
\end{algorithm}
The global spherical sampling algorithm (Alg.~\ref{alg:ibnss}) can thus be
sketched as follows. For a mesh $\mesh$ homeomorphic to the sphere, we first
construct $\psi$ and  the global mesh layout $\mesh_G$ on $\Sp^2$. We then construct
the target density $\nu_G$ by uniformly sampling $\mesh$ with a large
number of samples $m$ (importance sampling of the triangles from the face areas), and
projecting the samples onto $\mesh_G$. Note that $\nu_G$ is not
uniform on the sphere since it captures the
distortion induced by $\psi$. Finally, we use the NESOTS algorithm
to compute the sliced optimal transport sampling  $\mu_G$ and pullback this
measure onto the input mesh as described in Sec.~\ref{sec:implem}. 

\subsection{Local hyperbolic embedding}
\label{sec:local-hyperb-embedd}

If $\mesh$ has higher genus, a conformal map exists from $\mesh$ to
$\Hy^2$. Conformal coefficients can be obtained using the hyperbolic Discrete
Yamabe Flow formulation
\cite{luo2004combinatorial,bobenko2015discrete}. Please refer to 
Section \ref{sec:implem} for numerical details. The Yamabe flow
allows us to compute the per vertex conformal \dav{factors $\{u_i\}$, and then
the associated \dav{(hyperbilic})} edge length $l'_{ij}$ of the embedded mesh $\mesh_G$
onto $\Hy^2$.
\dav{From the updated metric, one can embed $\mesh_G$ onto the hyperboloid
of the Lorentz model (see Fig~\ref{fig:pipeline}) using a greedy
approach: starting from a initial vertex $V_0$ set to the origin $\xO$, triangles are layed down onto $\Hy^2$ in a greedy breadth first strategy process following Schmidt et al.'s approach \cite{schmidt2020inter}.}
If we continue the BFS visiting the triangles several times, this process reveals 
that the mapping from $\mesh$ to $\Hy^2$ is periodic and the conformal map {pave the entire hyperbolic plane}.
This prevents us from duplicating the global approach as described in Section~\ref{sec:sphere_sampl-alg}
since the image of the uniform measure $\mathcal{U}(\mesh)$ by a periodic function
is not integrable anymore, and hence the Optimal Transport framework cannot be used since 
it is only defined for probability measures.

\begin{figure}
  \begin{overpic}[width=8.5cm]{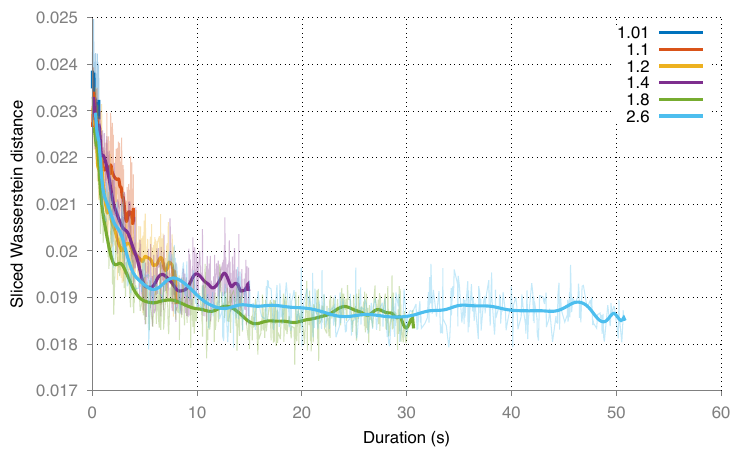}
    \put(17,43){\includegraphics[width=2cm]{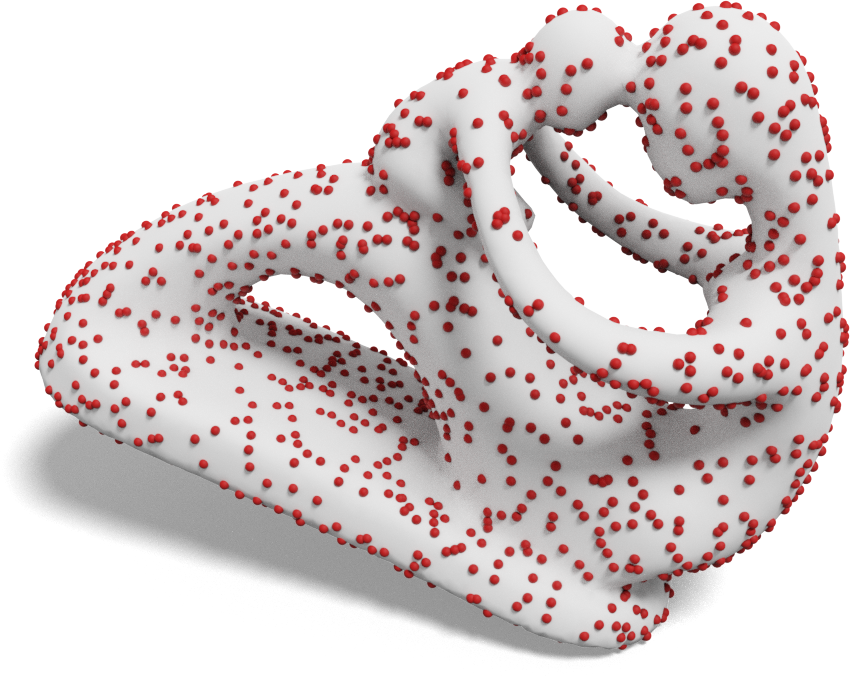}}
    \put(38,25){\includegraphics[width=2cm]{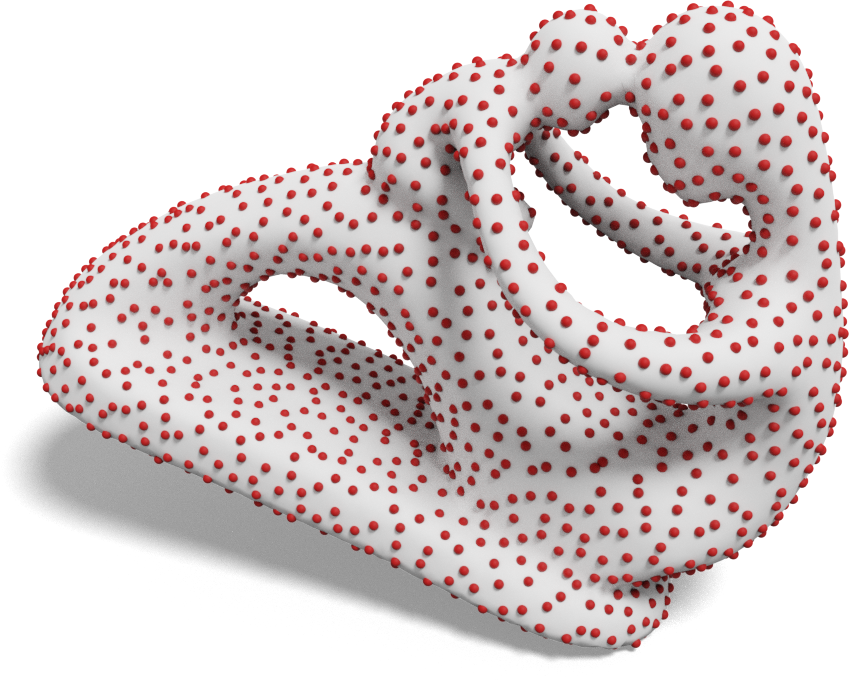}}
    \put(75,20){\includegraphics[width=2cm]{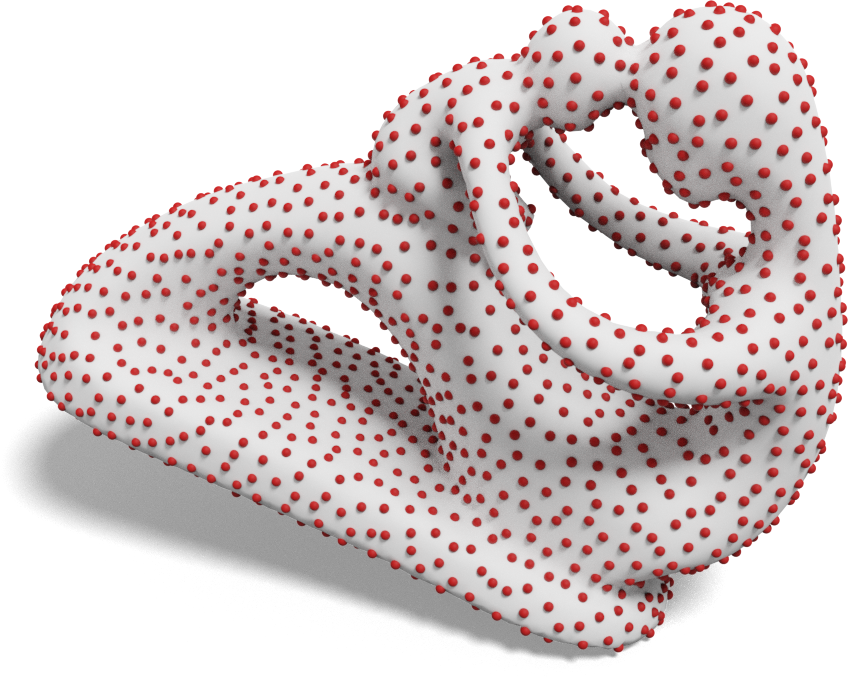}}
  \end{overpic}
  \caption{\textbf{NESOTS convergence results:} we illustrate the convergence of Alg.~\ref{alg:ilhbss} using $N=500$ iterations ($K=500$and $L=32$)
  for  2048 samples, as a function of the $\epsilon$ parameter. If $\epsilon$ is too small, local patches are small which implies short timing but low quality blue noise point pattern (as quantified by the SW distance to the uniform measure). As $\epsilon$ increases, the 
  blue noise quality is improved, but each iteration is longer. For $\epsilon\in\{1.01,1.1,1.2,1.4,1.8,2.6\}$, the average number 
  of $\mu_i$ samples  in each patch is respectively $\{3.31,14.76,29.97,61.86,124.92,242.82\}$. Sampled meshes correspond to the final step of $\epsilon\in\{1.1,1.4,2.6\}$
  respectively.\label{fig:convergence}}
\end{figure}

\dav{To overcome this problem we restrict the embedding to {patches} of the 
mesh (see Fig.~\ref{fig:pipeline} and Alg.~\ref{alg:ilhbss}): 
starting from a global Yamabe Flow that is  solved
only once, we iterate over a local layout construction with an  associated low distortion map $\psi_i$, and use NESOTS on this compact subset of $\Hy_2$.}
In this process, the choice of the first vertex of the layout matters
since the distortion will be very low in a neighborhood of $V_0$ (mapped to $\xO$), and will grow exponentially with the 
distance to it. Hence, using the embedding for $\Hy^2$ in $\mathbb{R}^3$, the main idea of the local algorithm is to construct a local layout until the (Euclidean) distance 
to the origin $\xO$,in the $z$ direction, exceeds a certain threshold $\epsilon$. 
As we will ignore 
triangles far from the origin, we only build low distortion mappings.
Note that the size of the patch for which the distortion is low depends on the 
quality of the mesh (triangle aspect), and on the curvature around $V_0$. 
The choice of $\epsilon$ allows controlling the scale of the optimization, 
giving a tradeoff between the sliced energy quality and speed \dav{(smaller patches leads to faster iterations)}.
The effect of $\epsilon$ is 
evaluated in Fig.~\ref{fig:convergence}. 
\begin{figure*}
  \centering
  \raisebox{0.3cm}{\begin{overpic}[width=12.5cm]{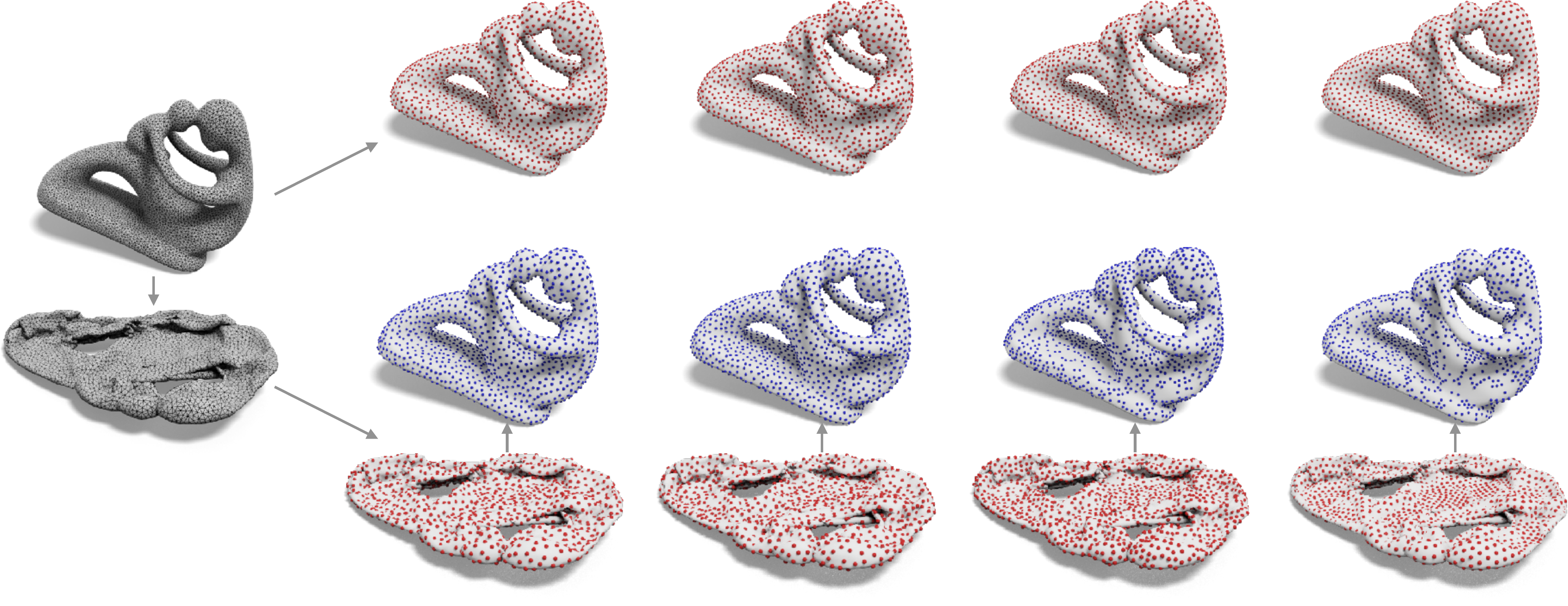}
    \put(9,-2.5){$(a)$}
    \put(30,-2.5){$(b)$}
    \put(50,-2.5){$(c)$}
    \put(70,-2.5){$(d)$}
    \put(91,-2.5){$(e)$}
    \put(121,-2.5){$(f)$}
   \end{overpic}}
   \raisebox{0.3cm}{\begin{overpic}[width=5cm]{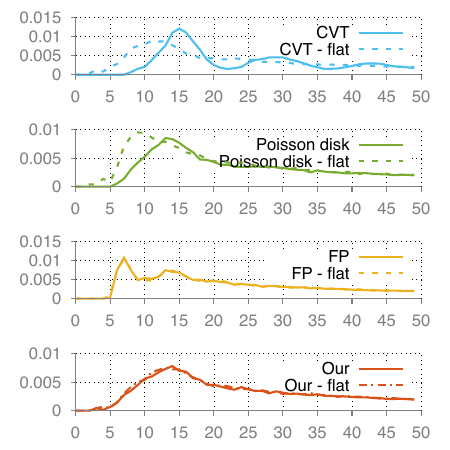}\put(45,-1){\color{gray}\small distances}\end{overpic}}
  \caption{\textbf{Intrinsic blue noise sampling of manifolds:} \dav{Given the \texttt{fertility} shape with two different Euclidean embeddings $(a)$. The flattened one is obtained through a physical simulation such that the two embeddings are intrinsically isometric}. We illustrate the sampling of the meshes with red dots using our approach $(b)$, the intrinsic farthest point approach (FP) \cite{peyre2006geodesic} $(c)$, the Poisson disk sampling in ambient space $(d)$, and
   the CVT sampling\cite{liu2009centroidal} $(e)$. The blue dots correspond to the sampling on the \emph{flat} embedding that are mapped to the 
  \emph{unflattened} one. First, we observe that our purely intrinsic approach leads to similar point sets in blue and red in $(c)$. 
  Best point patterns are obtained using CVT when the embedding is 
  correct in $\mathbb{R}^3$, \emph{i.e.} no thin layers ($(d)-$top). However, for both Poisson disk and CVT, the sampling of the \emph{flat} embedding leads to \dav{defective} 
  point patterns (holes in blue samples in $(d)$ and $(e)$). In $(f)$, we present pair correlation functions for each sampler (both on the flat and top row meshes).\label{fig:comparisons}}
  \batou{}
\end{figure*}

When a sample is displaced outside of the patch layout on $\Hy^2$, we just ignore the displacement (similarly to \cite{paulin2020sliced} when sampling $[0,1)^d$ or the d-Ball). 
  To make sure that all the points are optimized as equally as possible, we just keep track of the 
number of times a given vertex $\mesh$ has been used as the origin $v_0$ of a patch and iterate on the local patch construction starting by the least embedded vertex \dav{(the priority queue in Alg.~\ref{alg:ilhbss})}.
 Note that the local layout construction is extremely fast (linear
complexity in the number of triangles of the patch).

\begin{algorithm}\small
\caption{Intrinsic local hyperbolic blue noise surface sampling}
\label{alg:ilhbss}
\KwData{$\mesh$, $\nu$ on $\mesh$, $n$, $N$ , $K$, $L$, $\gamma$, $G=\Hy^2$ (see Alg.~\ref{alg:nesots})}
$\{u_i\} = \text{YamabeFlow}(\mesh)$ \;
${\nu}_G = \text{sampleMeshFaces}(\mesh_G,\nu,m)$ \;
\For{$i \in [[1,N]]$}{
    $vert = \text{PopVertexVisitPriorityQueue}()$\;
    $(V_i,F_i) = $ ComputeLocalHyperbolicLayout$(\{u_i\},vert,\epsilon)$ \;
    $\text{UpdateVertexVisitPriorityQueue}(V_i)$\;
    $\mu_i = $ ComputeRestrictionToLayout$(\mu, F_i)$\;
    $\nu_i = $ ComputeRestrictionToLayout$(\nu, F_i)$\;
    $\mu_G = \text{NESOTS}(\mu_i,\nu_i,K,L,\gamma)$ \tcp*{Alg.~\ref{alg:nesots}}
    $\mu = $ MapToMesh$(\mu_G,\mesh,\mesh_G)$  \tcp*{Alg.~\ref{alg:mmbm}}
}
\Return{$\mu$}
\end{algorithm}

\begin{figure*}[!htp]
  \subfigure[]{\includegraphics[width=4cm]{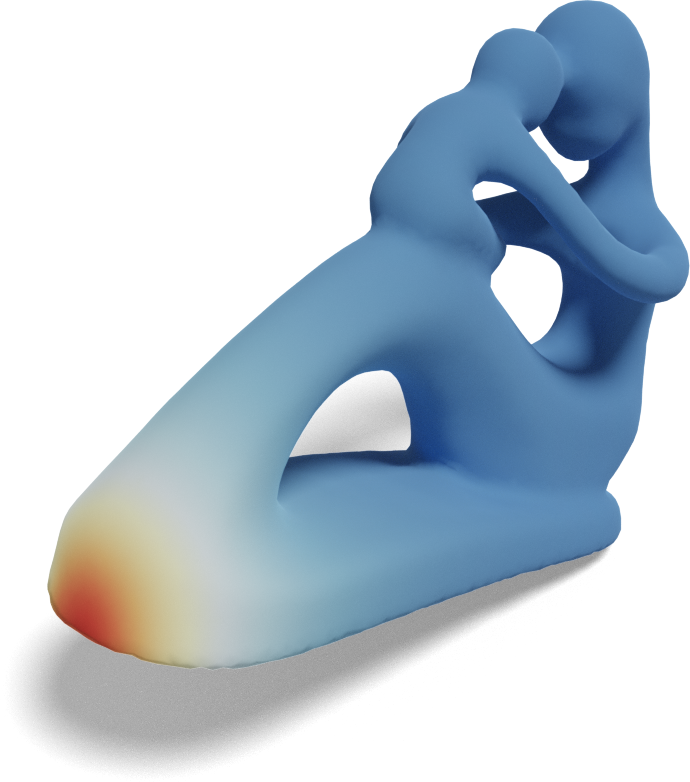}}
  \subfigure[]{\includegraphics[width=4cm]{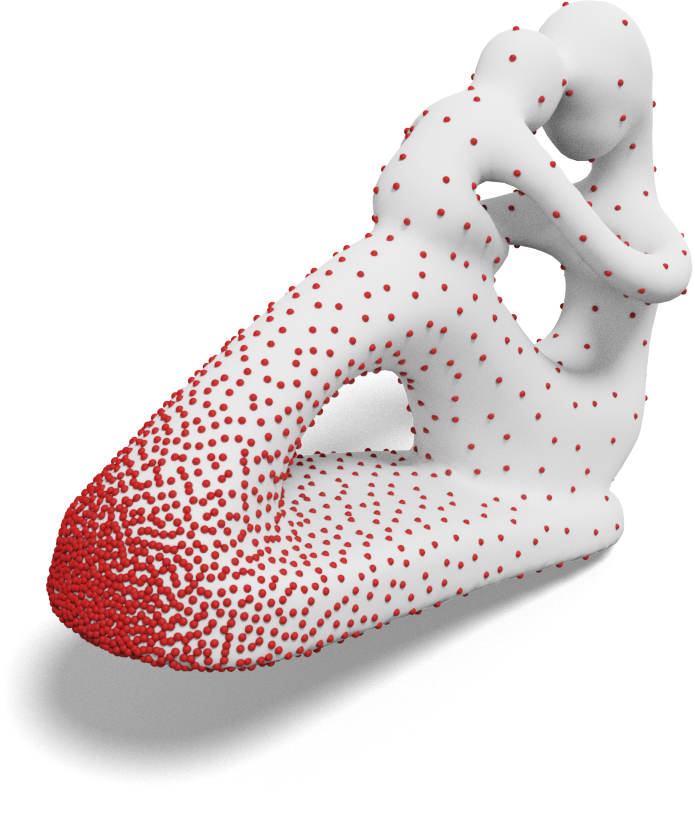}}
  \subfigure[]{\raisebox{.8cm}{\includegraphics[width=4cm]{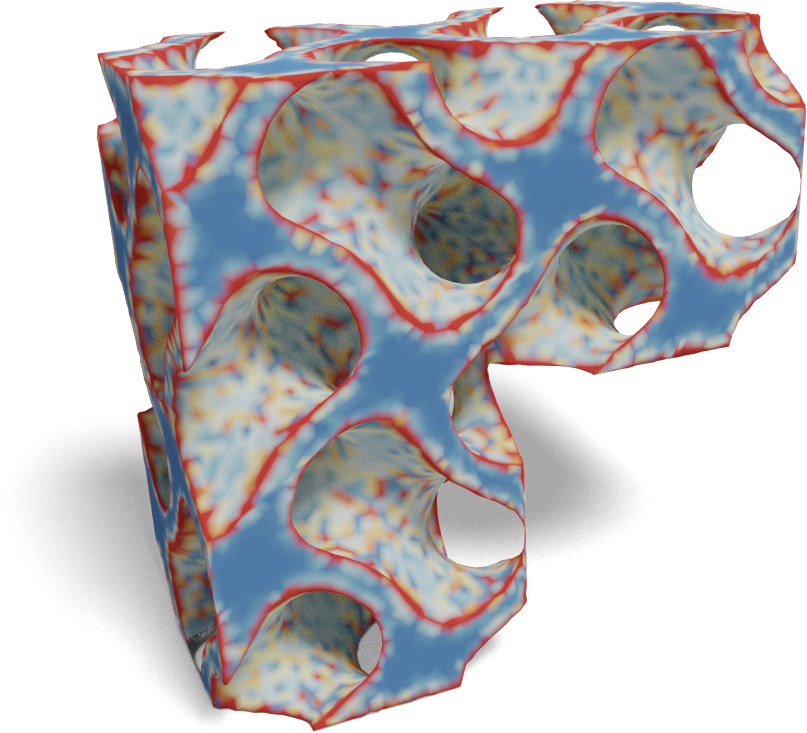}}}
  \subfigure[]{\raisebox{.8cm}{\begin{overpic}[width=4cm]{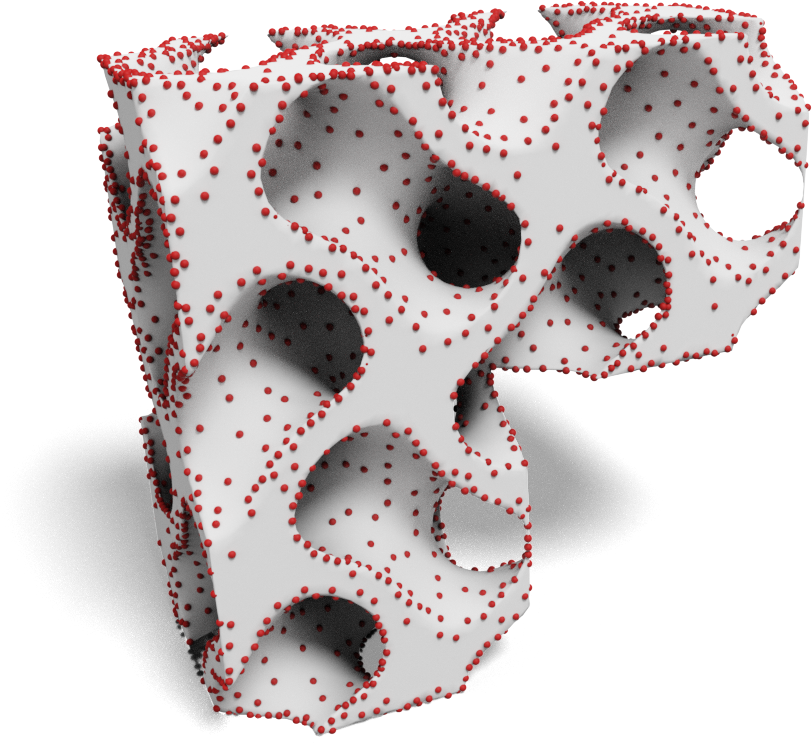}
    \put(70,-20){{\includegraphics[width=3cm]{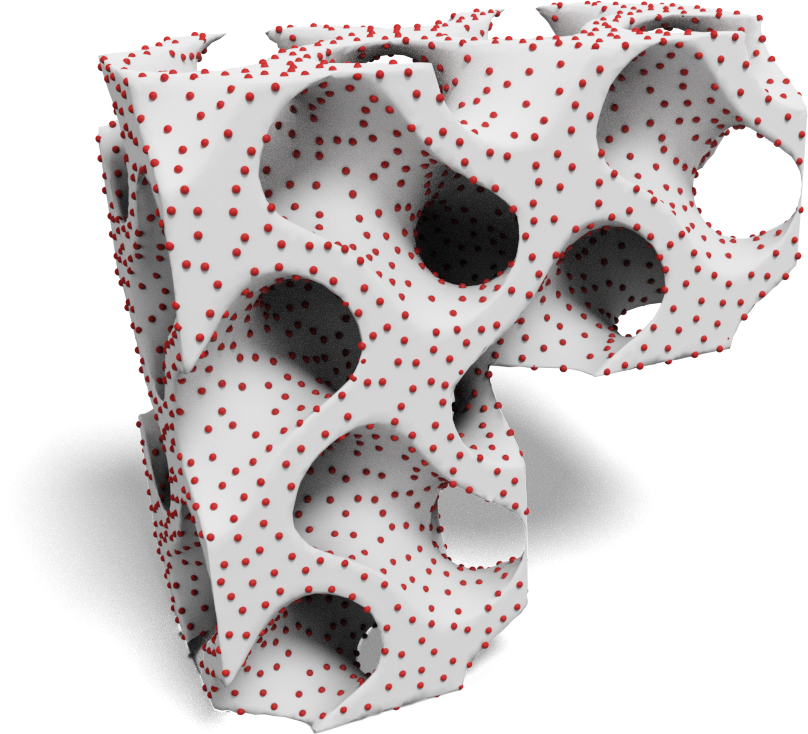}}}
  \end{overpic}}}
  \caption{\textbf{Non-uniform target density examples on meshes}: given an input probability density function, a smooth one $(a)$ on the \texttt{fertility} 
  shape (genus$-4$ manifold, AIM@shape) and mean curvature driven one in $(c)$ (gryroid surface, genus$-32$ manifold), 
  our approach is able to generate blue noise samples $\mu$ approximating the density (2048 samples for $(b)$ and 4096 samples for $(d)$). In $(d)$ we also 
  illustrate the sampling of the gyroid targeting  the uniform density for comparison.\label{fig:nonunif2}}
  \end{figure*}

\begin{figure*}\centering
  {\includegraphics[width=3cm]{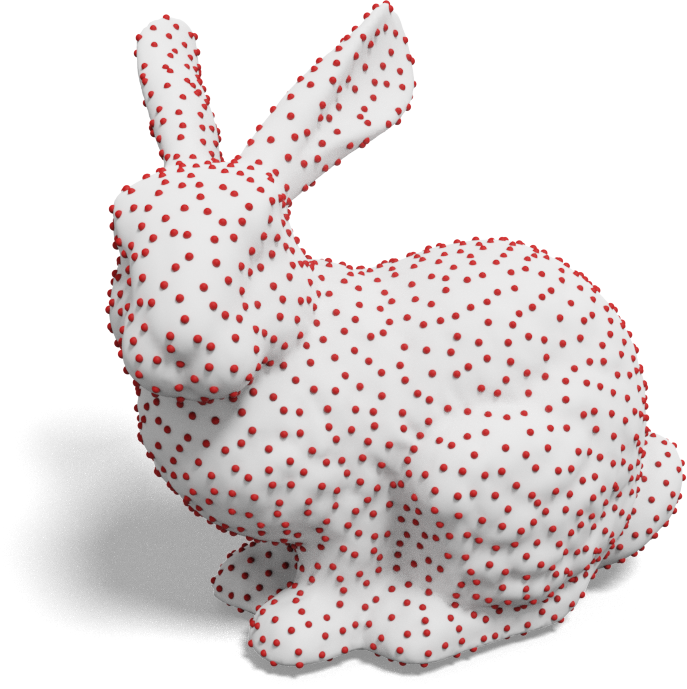}}
  \includegraphics[width=3cm]{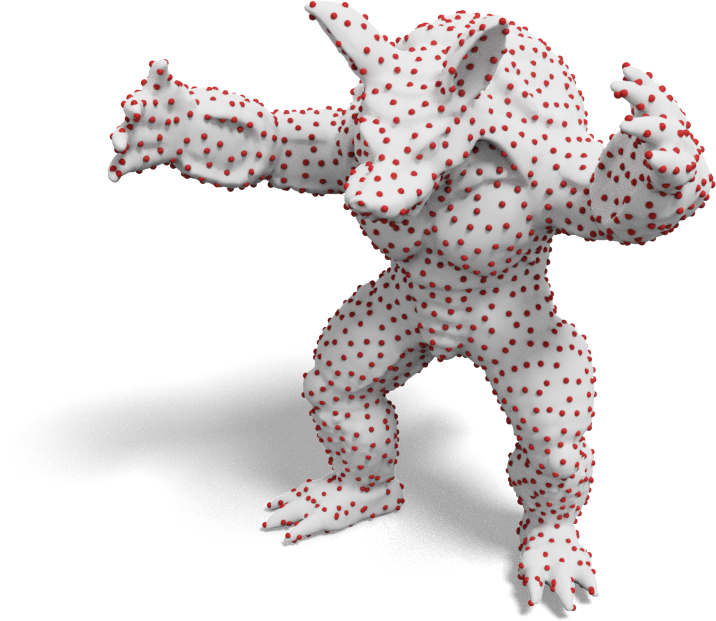}
  \raisebox{0.cm}{\includegraphics[width=3.2cm]{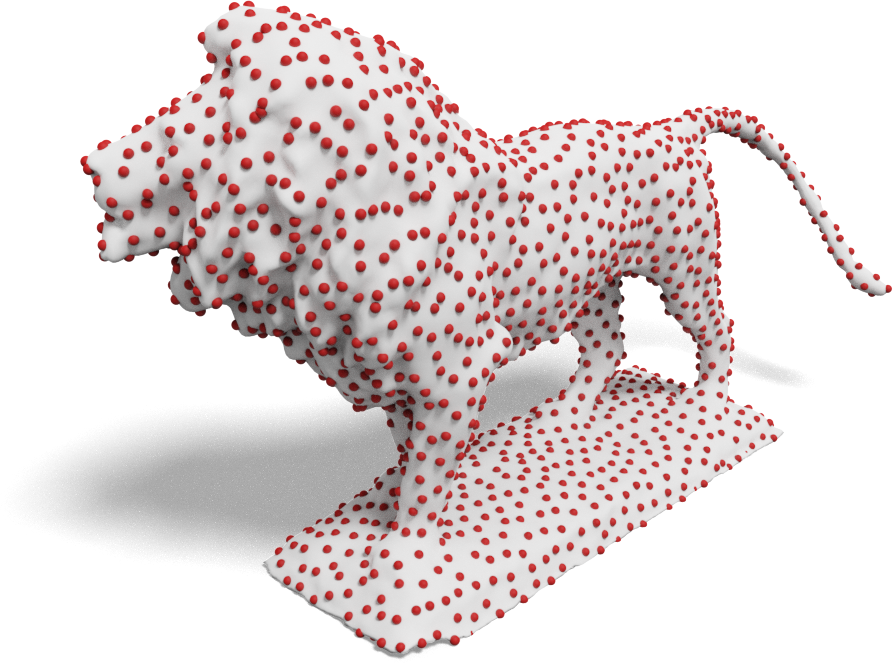}}
  \raisebox{-0.2cm}{\includegraphics[width=3cm]{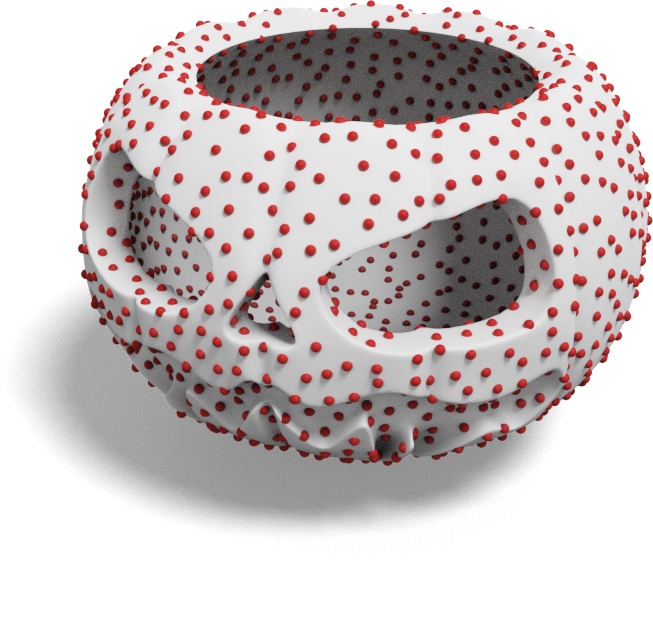}}
  \raisebox{-0.2cm}{\includegraphics[width=3cm]{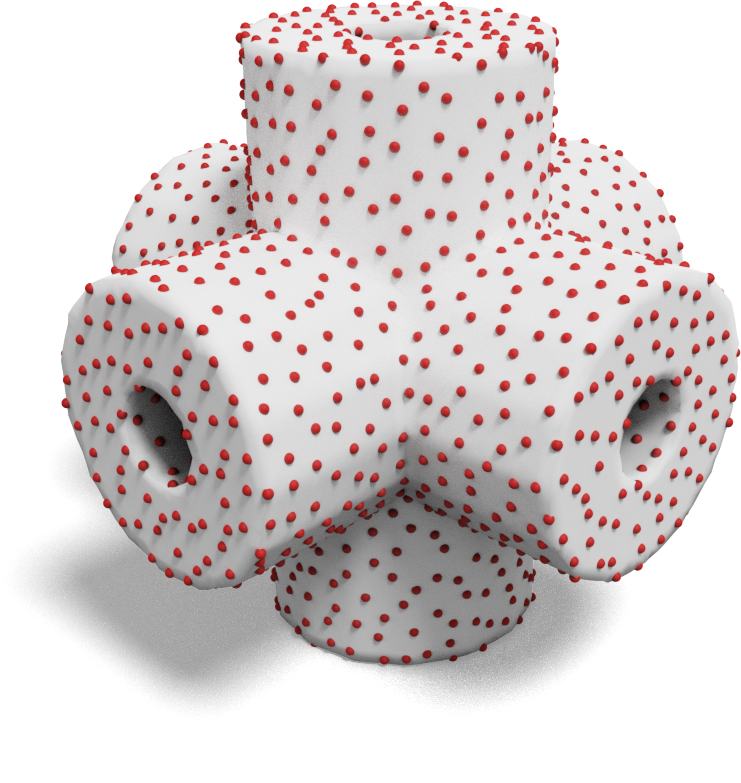}}
  \caption{\textbf{Intrinsic discrete manifold sampling:} additional sampling results with 2048 samples for $g=0$ and $g\geq 2$ surfaces.\label{fig:additional}}
\end{figure*}

In Fig.~\ref{fig:comparisons}, we demonstrate the interest of the intrinsic sampling on high genus meshes. When the embedding is ambient-compatible (first row), we observe a slightly better sample distribution using our approach than \dav{FP and }Poisson Disk sampling. In contrast, the CVT based approach produces a very high quality point pattern. Although, when the embedding is defective, our purely intrinsic approach led to an almost identical point pattern (in red) when mapped back to a better embedding (in blue) $(b)$, whereas both Poisson disk and CVT have critical voids and clusters due to bad assignments. To quantify this finding, we have computed the pair correlation function (pcf) \cite{illian2008statistical} the  exact geodesic distance on the manifold between each pair of samples \cite{mitchell1987discrete}. In Euclidean domains, pcf and radial mean power spectra capture similar point pattern characteristics \cite{singh2019analysis}. In Fig.~\ref{fig:comparisons}-$(f)$, we observe similar blue noise distribution (a peek at some characteristic distance and no too-close samples). We also observe that on the flat and non-flat meshes, our approach leads to similar pcfs. The pcfs CVT and Poisson disk are highly degraded on the flat geometries. 
In Fig.~\ref{fig:nonunif2} we present sampling examples of non-uniform target measures on meshes. Additional sampling results are  given in Fig.~\ref{fig:additional}.

\subsection{Implementation details and complexity}
\label{sec:implem}

First of all, for the hyperbolic case, discrete conformal coefficients $\{u_i\}$ are obtained by minimizing a convex 
energy, whose gradient and Hessian are given in \cite{bobenko2015discrete}. We thus apply a Newton descent approach with backtracking to ensure convergence. On the models presented in this paper, timings are detailed in Table~\ref{table:timings}. In the spherical case, we rely on the CEPS code provided by Gillespie et al. \cite{Gillespie:2021:DCE} to explicitly construct the spherical embedding. Once obtained, Alg.~\ref{alg:ibnss} is a direct application of Alg.~\ref{alg:nesots} with the same computational cost.

For the analysis of the local hyperbolic optimization (Alg.~\ref{alg:ilhbss}), we experimentally observe that the number of samples $\mu_i$ and $\nu_i$ on the layout grows linearly with $\epsilon$. If $C_\epsilon$ denotes the average computation cost per slice and per patch, using a batch size $L$, $K$ steps per patch and $N$ global iterations, we obtain a $\mathcal{O}(N\cdot K\cdot L \cdot C_\epsilon)$ complexity. Note that unless specified otherwise,  we have used $N = 500,K = 10, \epsilon = 1.5$ and  $L = 32$ for all experiments. Although performances were not our primary concern, typical timings are given in Table \ref{table:timings}.
Please refer to Appendix \ref{sec:weiszfelds-algorithm} for a discussion on the computational cost overhead when using the geometric median instead of simply averaging directions in  Alg.~\ref{alg:nesots}.

Once samples are optimized in, either  globally for $\Sp^2$, or locally for $\Hy^2$, we need an 
efficient way to retrieve the face of the mesh a given sample falls in (and the barycentric coordinates of that sample in the face). For that purpose, we exploit the convexity
of the domains: we first construct a BVH of the spherical or hyperbolic layout triangles and get the face id by shooting a ray through the origin $(0,0,0)$ and the sample (see Alg.~\ref{alg:mmbm} in Appendix \ref{sec:addit-algor}).
Finally, in the hyperbolic case, to avoid having to map all the $m$ points of $\tilde{\nu}$
on each layout, for each slice, we only map the $n$ points that are subsampled from $\tilde{\nu}$. \dav{Source code is available at \dav{\url{https://github.com/baptiste-genest/NESOTS}}.}
\begin{table}
  \setlength{\tabcolsep}{1pt}\small
\centering\begin{tabular}{l|p{2.3cm}|ccc|p{1cm}|c}
  Shape & Credits &$|V|$& $|F|$& $g$& Yamabe Flow & NESOTS\\ 
  \hline
  \texttt{spot} &\cite{crane2013robust} & 2930& 5856& 0& n.a. & 17.48  \\
  \texttt{duck} &deriv. of K. Crane & 29999& 60006& 3& 10.67 & 27.73 \\
  \texttt{fertility}& AIM@Shape  & 8192& 16396& 4& 3.02 & 15.13 \\
\texttt{macaca} & \cite{wiley2005evolutionary}&3494& 7000& 4 & 1.36 & 11,12 \\
\texttt{gyroid}&  Thingi10k \#111246&22115& 44354& 32& 30.37 & 1.94
\end{tabular}
\caption{\textbf{Timings.} Mesh statistics and typical timings (in seconds) for the $g\geq 2$ shapes using the parameters presented in Sec.~\ref{sec:implem} (AMD Ryzen 5000-H, 16 cores). \label{table:timings}}
\end{table}

\section{Real projective plane $\mathbb{P}^d$ sampling}
\label{sec:appl-high-dimens}

Many objects generated by vectors are defined regardless of their length
or  sign. For instance, the orthographic projection of a 3d shape in
the direction $\dpp$ is the same for all  $\lambda\dpp,  \forall \lambda
\neq 0$.  The space where collinear vectors are identified is called the
Projective Plane $\Pp^d$. One idea might be to project the points on
the sphere, which will successfully identify the vectors equivalent up
to a positive scale $\lambda > 0$ but not up to a sign. Hence, trying to
generate a "uniform" set of lines with any blue noise sampler on the
sphere does not output satisfactory results as the points are not
optimized to take into account this equivalence relationship. A simple
modification of Alg.~\ref{alg:nesots} described in Alg.~\ref{alg:pps},
allows us to successfully extend the blue noise generation of points, in any
dimension on $\Pp^d$ following any density on the sphere satisfying
$f(\xp) = f(-\xp)$ for $\xp \in \Sp^d$. To the best of our knowledge, this is new.

\begin{figure}
  \includegraphics[width=8cm]{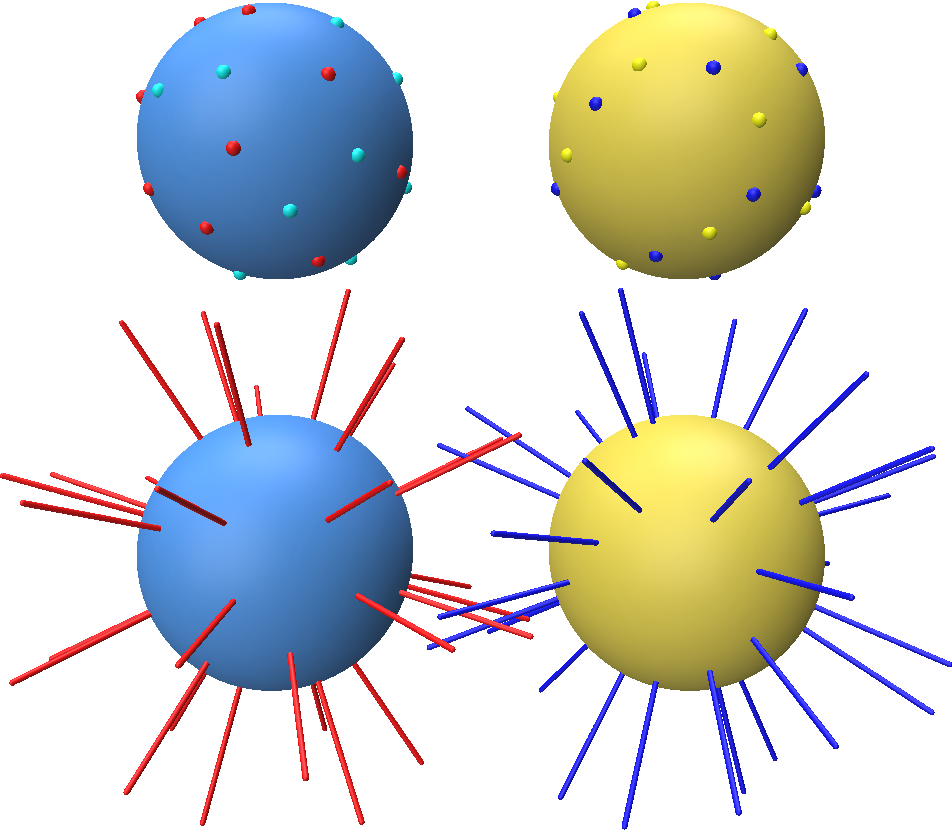}
  \label{fig:linesampling}
  \caption{\textbf{Projective plane $\Pp^2$ sampling:} red points are sampled with Alg.~\ref{alg:pps}, light blue
  points are the opposites of the red ones. \dav{Similarly, blue and yellow points are given by a spherical Fibonacci \cite{keinert2015spherical}. Points obtained }by Alg.~\ref{alg:pps} have \dav{better}  blue noise characteristics when considered with their opposites. To illustrate its use, we display at the bottom row the lines generated by the points. }
\end{figure}

\paragraph*{Lines and Hyperplanes sampling.}
As already stated, lines, characterized by their unit vector, can be generated uniformly on $\Pp^d$ using Alg.~\ref*{alg:pps}  (see Fig.~\ref{fig:linesampling} for a  3d  blue noise line sampling in $\mathbb{P}^2$). By taking the   orthogonal
complement of such lines, we can similarly obtain a blue noise sampling of $(d-1)-$hyperplanes.

{
\paragraph*{Affine line and hyperplane sampling.}
Note than even \emph{affine} spaces can be sampled by Alg.~\ref{alg:pps}. 
For instance, an affine line can described by its Cartesian equation, i.e. in dimension 2
\begin{align}
\label{eq:cart}
  a\xp + b\yp + c = 0,
\end{align}
but notice that, $\forall k \neq 0$, if $\xp$ and $\yp$ are solutions of \eqref{eq:cart}, then $ka\xp + kb\yp + kc = 0$. Hence each affine space of dimension $d$ can be represented in the projective plane $\Pp^d$ by 
its \dav{Cartesian} coefficients (here $(a,b,c)^t$). See Fig.~\ref{fig:affinelinesampling} for a 2d affine line sampling experiment.

\begin{figure}
  \begin{center}
  \includegraphics[width=3cm]{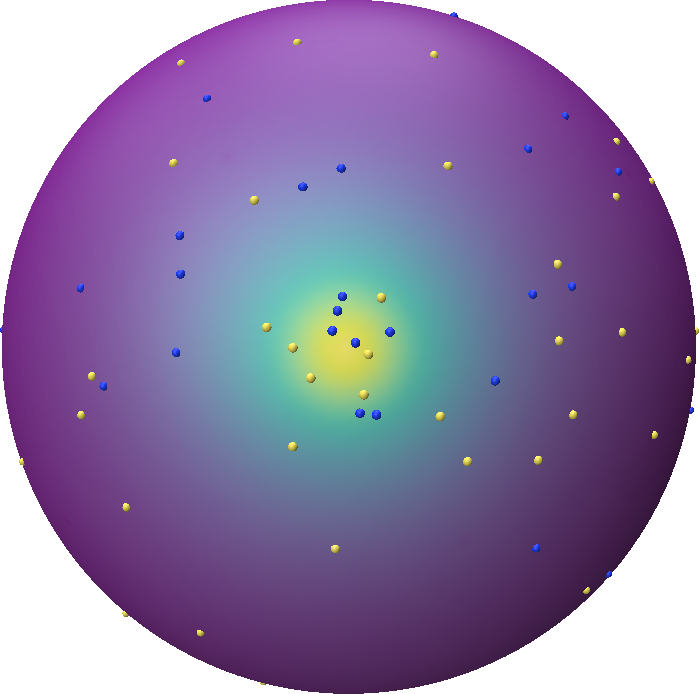}~~~~
  \includegraphics[width=3cm]{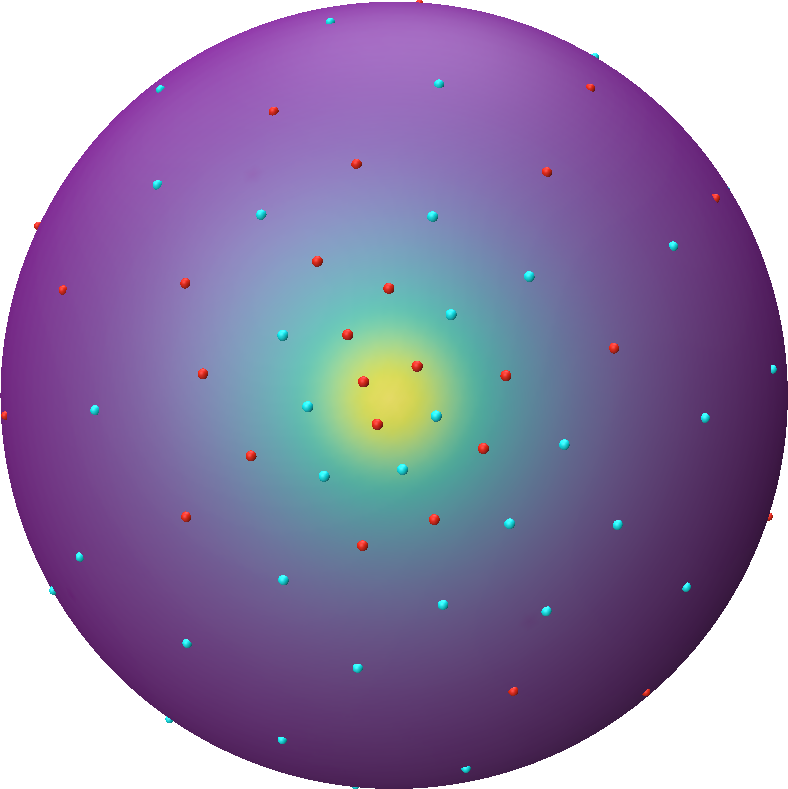}\\
  ~\\
  \includegraphics[width=3cm]{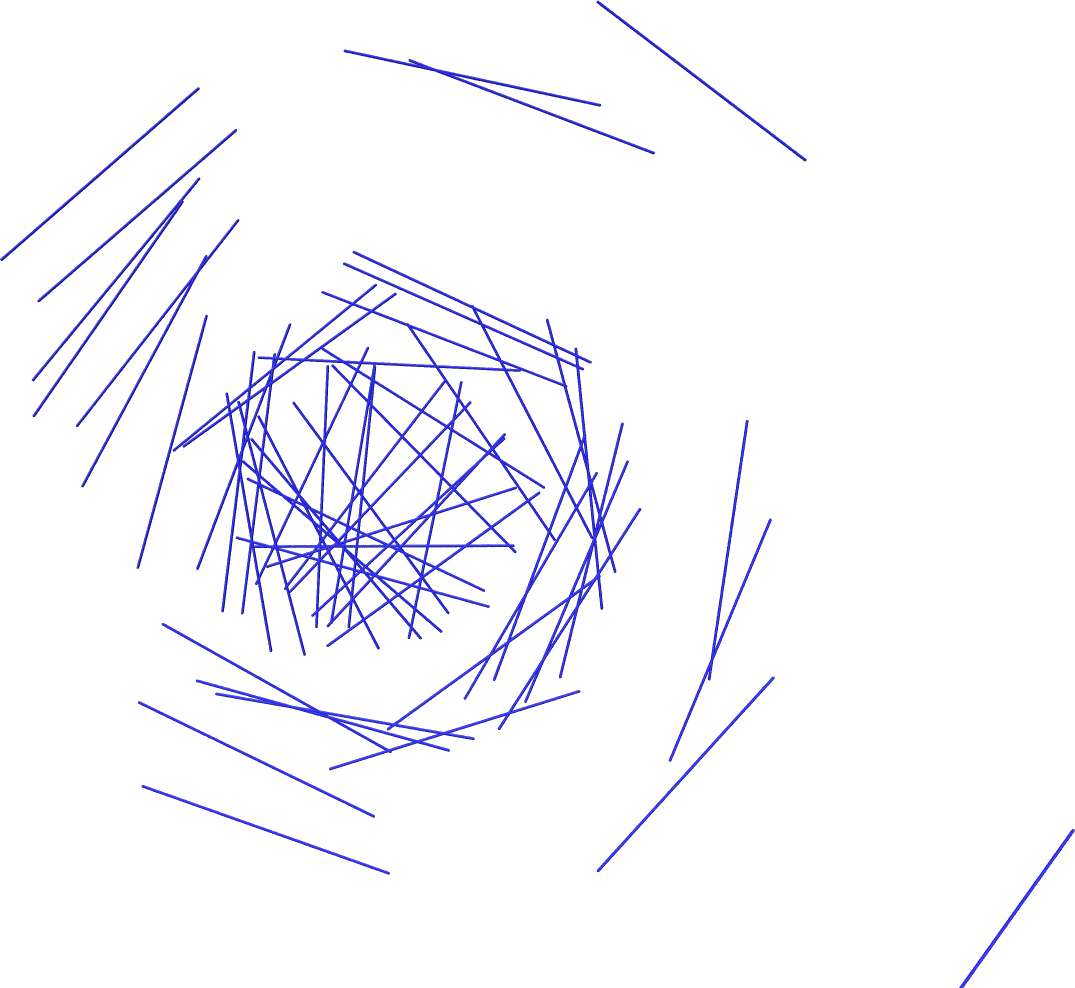}~~~~
  \includegraphics[width=3cm]{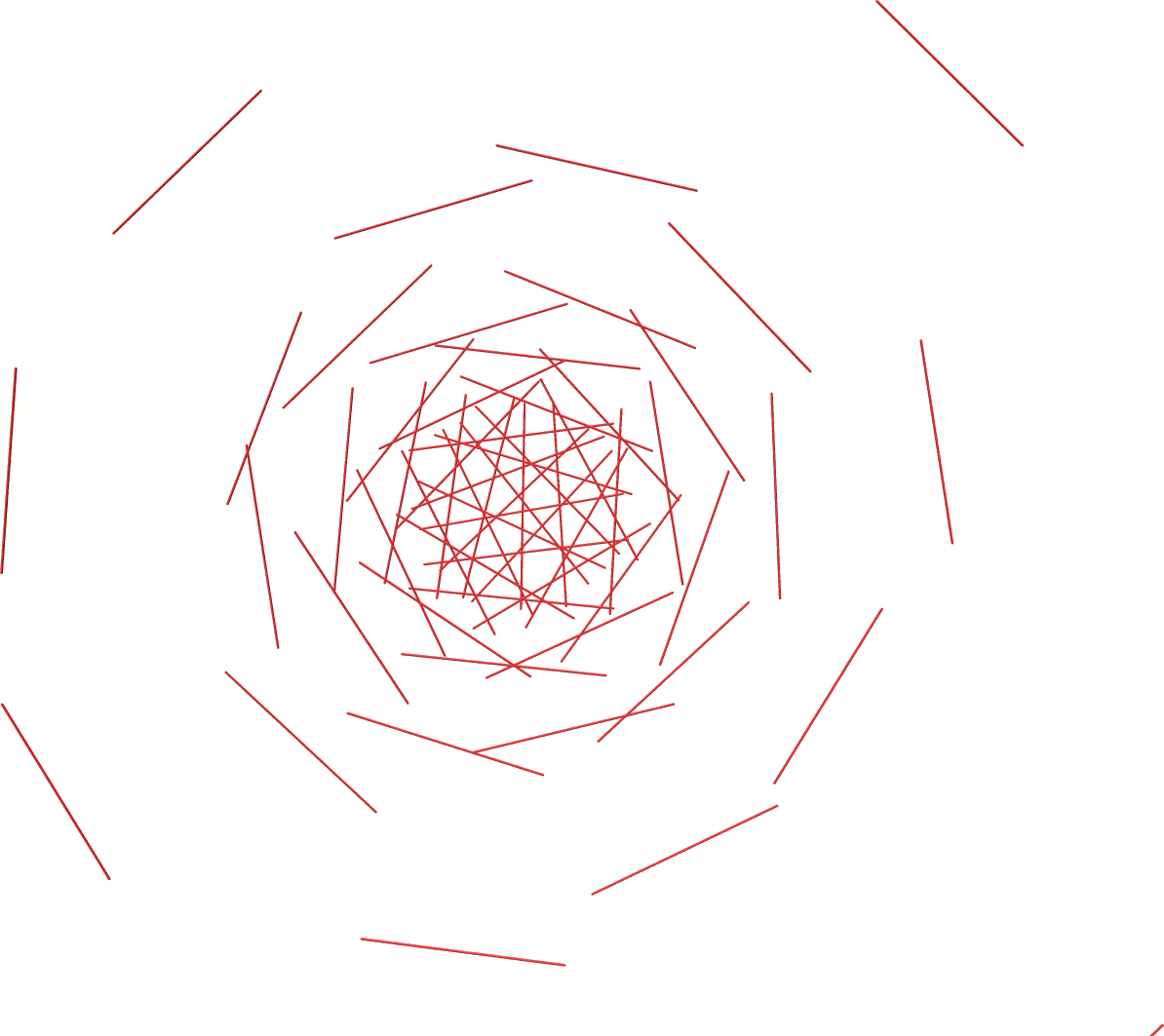}
  \end{center}
  \caption{\textbf{Affine lines sampling:} from the mapping of lines coefficients to $\Pp^2$, we generate 64 blue noise affine lines following a non-uniform density (top row)  using either a white noise sampling (left column) or Alg.~\ref{alg:pps}. When mapped back to $\Euc^2$, our sampling  exhibits blue noise characteristics in $\Euc^2$ with respect to the metric induced by the \dav{Cartesian} mapping (second row). Note that here only segments are displayed for the sake of clarity but they are actual lines of $\Euc^2$.  \label{fig:affinelinesampling}  }
\end{figure}
}

\paragraph*{Rotation Sampling by Unit Quaternion sampling.}
A unit quaternion $q$ can act on a vector as a rotation 
\begin{align*}
    \xp\mapsto \qp^{-1}\tilde{\xp}\qp\,,
\end{align*}
where $\tilde{\xp}$ is the imaginary quaternion with $\xp$ as vector
part. Since $\qp$ appears twice in the product, $\qp$ and $-\qp$ gives the
same rotation. Hence one can use Alg.~\ref{alg:pps} on $\Pp^3$ to
uniformize a set of unit quaternions (represented as unit 4
dimensional unit vectors). Previous approaches such as
\dav{Alexa's technique} \cite{alexa2022super},  provides good sampling on the 3-Sphere but
does not directly tackle the sign equivalence problem, which leads to
imperfect rotation sampling . The results of the rotation sampling
process is displayed in Fig.~\ref{fig:teaser}-\emph{(right)} where each shape is rotated by a
rotation generated by Alg.~\ref{alg:pps}.

\begin{algorithm}\small
\caption{Real Projective Plane Sampling $\mathbb{P}^d$}
\label{alg:pps}
\SetKwFor{ParallelFor}{parallel for}{do}{end}
\KwData{$\nu = \sum_{i=1}^m \delta_{\yp_i}$, $K$, $L$, and $\gamma$ (see Alg.~\ref{alg:nesots}).}
\KwResult{$\mu^{(K)}$}

$\mu^{(0)}= \text{subSample}({\nu},2n)$\\
\For{$ j \in [[1,K]]$}{
\ParallelFor(\tcp*[f]{Batch}){$ l \in [[1,L]]$}{
    $\tilde{\nu} = \text{subSample}(\tilde{\nu},2n)$ \tcp*{Sec.~\ref{sec:non-unif-dens}}
    $\theta$ = RandomSlice() \tcp*{Sec.~\ref{sec:geodesic-slices}}
    $\tilde{\nu}_\theta = P^\theta\left(\tilde{\nu}^l\right)$\tcp*{Sec.~\ref{sec:geodesic-slices}}
  $\mu_\theta = P^\theta\left(\mu^{(j)}\right) \cup -P^\theta\left(\mu^{(j)}\right)$\tcp*{Sec.~\ref{sec:geodesic-slices}}
  $T = \text{Solve1DOT}({\mu}_\theta$,$\tilde{\nu}_\theta)$ \tcp*{Sec. \ref{sec:sorting-along-slice}}
    \For{$ i \in [[1,2n]]$}{
        $\gp = \Gamma_\theta\left(P^\theta\left(\xp^{(j)}_i\right),T\left(P^\theta\left(\xp^{(j)}_i\right)\right)\right)$ \tcp*{Sec.~\ref{sec:trans-group-acti}}
        $\dpp^l_i = \text{Log}_{\xp^{(j)}_i}\left (\gp\left ( \xp^{(j)}_i\right)\right)$ \tcp*{Sec.~\ref{sec:exp-log-map}}
    }   
}
    \ParallelFor{$ i \in [[1,n]]$}{
      $\dpp_i = \text{GeoMed}\left(\{\dpp^l_i\}_L \cup \{-\dpp^l_{i+n}\}_L \right)$ \tcp*{Sec.~\ref{sec:geometric-median}}
    $\xp_i^{(j+1)} = \text{Exp}_{\xp^{(j)}_i}\left(\gamma\, \dpp_i\right)$ \tcp*{Sec.~\ref{sec:riem-grad-desc}}
    }   
}
\Return $\mu^{(K)}=\sum_{i=1}^m \delta_{\xp^{(K)}_i}$
\end{algorithm}

\section{Limitations and future Work}

Our approach extends the blue noise sampling of any probability measure through the 
sliced optimal transport energy, originally designed for Euclidean domains, to Riemannian manifolds: 
the spherical space $\Sp^d$, the hyperbolic space $\Hy^d$, and the projective one $\mathbb{P}^d$. In a nutshell, from explicit advection and direction averaging steps on these spaces, we present a gradient descent strategy that optimizes
a point set minimizing the sliced Wasserstein energy. 

First of all, concerning the generic NESOTS approach, there are many opportunities for performance improvements. We are convinced that 
many variance reduction techniques could be borrowed from Monte Carlo rendering approach to speed up the sliced strategy (\emph{e.g.} importance sampling of the $\theta$ directions, control variates using a proxy for the SW energy). 

Thanks to the uniformization theorem, we demonstrated the interest of the approach for intrinsic blue noise 
sampling of discrete surfaces. Although we may not compete with existing extremely fast restricted Voronoi  based techniques when
the mesh has a good embedding, we advocate that the purely intrinsic nature of our construction is of interest. An important limitation
is the \dav{robustness of the global conformal map in the spherical case that may impact the sampling when high distortion occurs.}
\dav{In the hyperbolic case, our} local construction mitigates  this by controlling potential distortion issues (the $\epsilon$ parameter) but we are convinced that improvements exist, \emph{e.g.} using implicit intrinsic remeshing as in \dav{Gillespie et al.} \cite{Gillespie:2021:DCE}.
On the geometric side, we only focused $g=0$ and $g\geq 2$ surfaces leaving the flat metric space case aside. For $g=1$, cut-and-open strategies must be designed that we avoid in spherical and hyperbolic domains. 
In this paper, we also focus on the sample generation, leaving the use cases of the point set as future work (\emph{e.g.} decal placement, function reconstruction, remeshing). 
\setlength{\columnsep}{5pt}%
\begin{wrapfigure}{r}{4cm}
  \vspace{-10pt}
  \includegraphics[width=4cm]{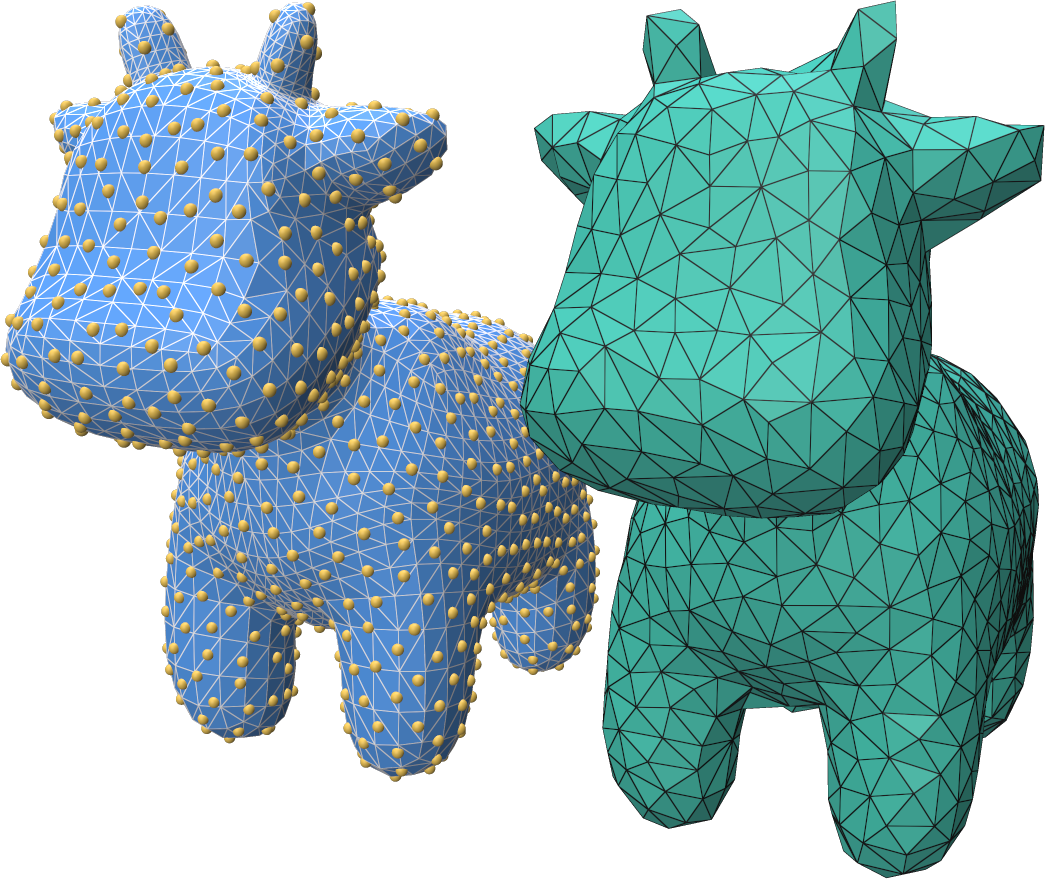}
  \vspace{-20pt}
\end{wrapfigure}
For remeshing, the convexity of the $\Sp^2$ and $\Hy^2$ could be further \dav{exploited} to reconstruct a mesh: on the $\Sp^2$ the convex hull of the optimized samples leads to a trivial (manifold) surface reconstruction (see inset). 
The hyperbolic case is more complicated as holes could be embedded in a compact subset of $\Hy^2$ for which the global convex hull topology does not make sense. We believe that a local combinatorial construction from the convex hull using a small $\epsilon$ could be investigated.

Finally, we have only scratched the use of blue noise sampling in the projective space $\mathbb{P}^d$ for computer graphics applications. For instance, Monte Carlo-like line and segment sample estimators may lead to drastic reductions of variance in rendering for some  effects such as soft shadows or defocusing blur \cite{singh17variance}. We believe that affine line sampling approaches as illustrated in Fig.~\ref{fig:affinelinesampling} would be of great interest in this context.

\section*{Acknowledgments}
This research was partially funded by the projects StableProxies (ANR-22-CE46-0006) and OTTOPIA (ANR-20-CHIA-0030) of the French National Research Agency (ANR),  and gifts from Adobe Inc.

\appendix
\section{Additional algorithms}
\label{sec:addit-algor}

{
The objective of Alg.~\ref{alg:mmbm} is to find the face a point is lying on, and to compute the correspondence between
its position on the face embedded in $\Euc^3$ and on the layout in $\Sp^2$ (resp $\Hy^2$) through barycentric coordinates.
Even if we theoretically should use spherical (resp. hyperbolic) barycentric coordinates, we observe that the Euclidean barycentric coordinates make a good enough quality proxy while avoiding computing transcendental functions at each mapping.
For high performances, the face retrieval can be done leveraging the convexity of $\Sp^2$ and $\Hy^2$ through a ray shooting  approach (rays starting from the domain origin to the sample to locate), with a  BVH of the faces. 
In our implementation, we used the library \cite{bvh}.
}
\begin{algorithm}
\caption{Mapping measures between two meshes}
\label{alg:mmbm}
\KwData{$\mu_G$ ,$\mesh$ and $\mesh_G$}
BVH $= \text{BuildBVH}(\mesh_G)$ \;
\For{$i \in [[1,n]]$}{
    $\tilde{F} = $ BVH.intersect$(\mesh_G,\xO,\xp^G_i)$ \;
    $b_i = $ ComputeBarycentricCoordinates$(M_G,\xp^G_i,\tilde{F})$ \;
    $F = \text{FindCorrespondingFace}(\tilde{F},\mesh)$ \;
    $\xp_i = $ PositionFromBarycentricCoordinates$(\mesh,b_i,F)$ \;
}
\Return{$\mu$}
\end{algorithm}
\label{sec:weiszfelds-algorithm}
In Alg.~\ref{alg:GM}, we detail the  Weiszfeld's algorithm we use for the
geometric median computation using an iterative least squares
approach. \batou{Note that, as stated in Section \ref{sec:geometric-median}, Weiszfeld's algorithm is used to combine the gradients (in $\mathbb{R}^n$) during the Riemannian stochastic gradient descent.}
\begin{algorithm}\small
\caption{Weiszfeld's geometric median algorithm \cite{weiszfeld1937point}}
\label{alg:GM}
\KwData{The samples $\{\xp_i\}_L\in\mathbb{R}^d$, a stability parameter  $\tau\in\mathbb{R}$}
$\yp = 0$ \;
$j = 2\tau$ \;
\While{$j > \tau$}{
    $d = 0$ \;
    $w = 0$ \;
    $\tilde{\yp} = 0$\;
    \For{$i \in [[0,L]]$}{
        $d = \tau + \|\yp-\xp_i\|_2$ \;
        $w \mathrel{{+}{=}} d$ \;
        $\tilde{\yp} \mathrel{{+}{=}} \frac{\xp_i}{d}$ \;
    }
    $\tilde{\yp} \mathrel{{/}{=}} w$ \;
    $j = \|\yp-next\|_2$ \;
    $\yp = \tilde{\yp}$ \;
}
\Return{$\yp$}
\end{algorithm}
{
Theoretically, without the $\tau$ term, this algorithm does not converge if $\yp_0 = \xp_i$ for some $i$. 
In practice, with $\tau > 0$, we do not observe convergence issues (interested readers may refer to \dav{Cohen et al.} \cite{cohen2016geometric} for a review of standard algorithms).
While geometric median is an essential element to guarantee
quality of the result for highly non-uniform density functions, a slight computational overhead exists when compared to the geometric mean. On the \texttt{fertility} mesh with standard parameters (see Sec~\ref{sec:implem}), the optimization part of the Alg.~\ref{alg:ilhbss} takes 12.38s with the mean and 13.33s with the geometric median ($L=32$ and $\tau=10^{-7}$ for all experiments). 
}

\bibliographystyle{eg-alpha-doi}
\bibliography{references}

\end{document}